\documentclass[12pt,twocolumn]{emulateapj}
\slugcomment{{\it Submitted to ApJ 09/30/08; Accepted 11/07/08}}

\begin{document}

\def\s{{\rm\,s}} 
\def\erg{{\rm\,erg}} 
\def\cm{{\rm\,cm}} 
\def\m{{\rm\,m}} 
\def\mum{\,\mu{\rm m}} 
\def\km{{\rm\,km}} 
\def\mm{{\rm\,mm}} 
\def\gm{{\rm\,g}} 
\def\g{{\rm\,g}} 
\def\kg{{\rm\,kg}} 
\def\au{{\rm AU}} 
\def\deg{{\rm deg}} 
\def\rad{{\rm rad}}   
\def\AU{{\rm\, AU}}  
\def\K{{\rm\,K}}  
\def\yr{{\rm\,yr}}  
\def\Hz{{\rm\,Hz}}  
\def\eV{{\rm\,eV}}  
\def\pomega{\widetilde{\omega}}
\def\baru{{\rm\,bar}}
\def\ion#1#2{#1$\;${\small\rm\@{#2}}\relax}
\def\beq{\begin{equation}}
\def\eeq{\end{equation}} 

\shortauthors{Chiang et al.}
\shorttitle{Fomalhaut's Disk and Planet}

\title{Fomalhaut's Debris Disk and Planet:\\Constraining the Mass of Fomalhaut b From Disk Morphology}

\author{E.~Chiang\altaffilmark{1,2}, E. Kite\altaffilmark{2}, P. Kalas\altaffilmark{1}, J. R. Graham\altaffilmark{1}, \& M. Clampin\altaffilmark{3}}

\altaffiltext{1}{Department of Astronomy,
601 Campbell Hall,
University of California at Berkeley,
Berkeley, CA~94720, USA}
\altaffiltext{2}{Department of Earth and Planetary Sciences,
307 McCone Hall,
University of California at Berkeley,
Berkeley, CA~94720, USA}
\altaffiltext{3}{Goddard Space Flight Center,
Greenbelt, MD~20771, USA}

\email{echiang@astro.berkeley.edu}

\begin{abstract}
  Following the optical imaging of exoplanet candidate Fomalhaut b
  (Fom b), we present a numerical model of how Fomalhaut's debris disk
  is gravitationally shaped by a single interior planet. The model is
  simple, adaptable to other debris disks, and can be extended to
  accommodate multiple planets.
  If Fom b is the dominant perturber of the belt, then to produce the
  observed disk morphology it must have a mass $M_{\rm pl} < 3 M_{\rm
    J}$, an orbital semimajor axis $a_{\rm pl} > 101.5 \AU$, and an
  orbital eccentricity $e_{\rm pl} = 0.11$--0.13. 
  If the planet's orbit is apsidally aligned with the belt's, our model
  predicts $M_{\rm pl} = 0.5 M_{\rm J}$, $a_{\rm pl} = 115 \AU$, and
  $e_{\rm pl} = 0.12$.
  These conclusions
  are independent of Fom b's photometry. To not disrupt the disk, a
  greater mass for Fom b demands a smaller orbit farther removed
  from the disk; thus, future astrometric measurement of Fom b's
  orbit, combined with our model of planet-disk interaction, can be
  used to determine the mass more precisely.  The inner edge of the
  debris disk at $a \approx 133\AU$ lies at the periphery of Fom b's
  chaotic zone, and the mean disk eccentricity of $e\approx 0.11$ is
  secularly forced by the planet, supporting predictions made prior to
  the discovery of Fom b. However, previous 
  mass constraints based on disk morphology rely on several
  oversimplifications.  We explain why our constraint is more
  reliable. It is based on a global model of the disk that is not
  restricted to the planet's chaotic zone boundary. Moreover, we
  screen disk parent bodies for dynamical stability over the system
  age of $\sim$100 Myr, and model them separately from their dust
  grain progeny; the latter's orbits are strongly affected by
  radiation pressure and their lifetimes are limited to $\sim$0.1 Myr
  by destructive grain-grain collisions.  The single planet model
  predicts that planet and disk orbits be apsidally aligned.
  Fomalhaut b's nominal space velocity does not bear this out, but the
  astrometric uncertainties may be large.  If the apsidal
  misalignment proves real, our calculated upper mass limit of 3
  $M_{\rm J}$ still holds. 
  Parent bodies are evacuated from mean-motion resonances with Fom b;
  these empty resonances are akin to the Kirkwood gaps opened by Jupiter.
  The belt contains at least $3 M_{\oplus}$ of solids
  that are grinding down to dust,
  their velocity dispersions stirred
  so strongly by Fom b that collisions are destructive.
  Such a large mass in solids is
  consistent with Fom b having formed {\it in situ}.
  \end{abstract}


\section{INTRODUCTION}\label{sec_intro}

A common proper motion companion to Fomalhaut has been imaged by Kalas
et al. (2008, hereafter K08) using the Hubble Space
Telescope Advanced Camera for Surveys (HST ACS) coronagraph. Fomalhaut
b (Fom b) orbits interior to the system's well-known circumstellar
belt of dust \citep[e.g.,][hereafter K05, and references
therein]{holland98,kalas05}. 
While Fom b's
ultra-low luminosity leaves little doubt that it is of remarkably low
mass, sitting well below the regime of brown dwarfs, the
question remains: Just how low is its mass?

Based on the observed broadband spectrum of Fom b, K08 estimate an
upper limit on the mass $M_{\rm pl}$ of about $3 M_{\rm J}$. We
recapitulate their reasoning as follows.  Fom b is detected in HST's
F814W (0.7--0.9 $\mu$m) and F606W (0.45--0.7 $\mu$m) passbands in 2006. The
F606W flux is variable; the flux in 2006 was about half
that in 2004. Observations in 2005 with Keck in H band
(1.5--1.8 $\mu$m) and in 2008 with Gemini in L-prime (3.2--4 $\mu$m)
give only upper limits.

The F814W flux (observed only in 2006; imaging was not attempted
in 2004 for this passband) can be reproduced by thermal
emission from a 2--$4 M_{\rm J}$, 200-Myr-old planet that formed by
core accretion and is of supersolar metallicity
\citep{hub05,fortney08}.  Unfortunately, this same model atmosphere
underpredicts, by more than an order of magnitude, the F606W
fluxes. At the same time it overpredicts the H-band $3\sigma$ upper limit by a
factor of a few, and is marginally consistent with the L-prime $3\sigma$
upper limit. From considerations outlined by
K08, we can construct two, not entirely exclusive hypotheses.
Hypothesis one: F814W still traces planetary thermal emission, F606W
is contaminated by variable H$\alpha$ emission, and the atmospheric model
requires revision in H band.  The H$\alpha$ hypothesis is
inspired by variable H$\alpha$ emission from chromospherically active
and/or accreting low-mass stars and brown dwarfs.  The puzzling brown
dwarf GQ Lup B offers a possible precedent \citep{marois07}; in either
the case of Fom b or that of GQ Lup B, the H$\alpha$ luminosity
necessary to explain the anomalously large F606W flux is $\sim$1\%
that of the bolometric luminosity (unfortunately in neither case has
an optical spectrum been taken). As for uncertainties in model
exoplanet atmospheres \citep{fortney08,bsl03}, these appear greater in
H band than in F814W; the various models disagree with each
other at near-infrared wavelengths by factors of a few.  Hypothesis
two: both F814W and F606W are contaminated by starlight reflected off
an optically thick circumplanetary disk, and the F606W variability
arises from variable disk accretion onto the planet, possibly
implicating H$\alpha$ again.  To explain the detected fluxes in 2006
using reflected light alone, such a disk
would have to be comparable in size to the orbits of Jupiter's
Galilean satellites. It is not clear, however, how such a disk would
survive for the system age of $\sim$100 Myr; protosatellite disks are
thought to evolve on timescales shorter than a few Myr
\citep[e.g.,][]{canupward,mosque}.

Since both hypotheses admit the possibility of additional sources
of luminosity apart from planetary thermal emission, the 
$3 M_{\rm J}$ inferred from the observed F814W flux should be considered
an upper limit. As a whole, this interpretation
of Fom b's photometry is preliminary,
subject to significant revision as both observations and theory improve.

An alternative route to probing the properties of Fom b
is to exploit the morphology of the circumstellar belt.
Prior to the discovery of Fom b,
this approach was taken by \citet[][hereafter Q06]{quillen06}, who built
on earlier work by \citet{wyatt99}.
The striking intrinsic ellipticity of the belt of $e \approx 0.11$
can be forced by secular gravitational interaction with a single planet
on a similarly eccentric orbit interior to the belt.\footnote{See
the textbook by \citet{md00} for an introduction to secular perturbation
theory, which essentially treats masses as orbit-averaged elliptical wires.}
Fomalhaut b was predicted by Q06 to occupy an orbit
of eccentricity 0.1 and semimajor axis $\sim$119 AU, and to have
a mass between about $2\times 10^{-5}$ and $7\times 10^{-5}$ that of the
central star. For a stellar mass of $M_{\ast} = 2.3 M_{\odot}$,
this range corresponds to $\sim$0.05--0.2 $M_{\rm J}$.

The predictions of Q06 rest on the idea that the belt inner edge,
at semimajor axis $a_{\rm inner} = 133\AU$, is located at the outer
boundary of Fom b's chaotic zone. The chaotic zone is a swath of space
enclosing the planet's orbit in which test particle orbits are
chaotic and short-lived, as a consequence of overlapping first-order
mean-motion resonances \citep{w80}. The semimajor axes $a_{\rm
  chaotic}$ that bound this chaotic zone are displaced to either side of
the planet's semimajor axis $a_{\rm pl}$ by \beq \Delta a_{\rm chaotic} =
\left| a_{\rm chaotic} - a_{\rm pl} \right| \approx 1.3
\,\mu^{2/7} \label{eq_27} \eeq where $\mu = M_{\rm pl}/M_{\ast}$. The
coefficient of 1.3 arises from Wisdom's approximate scaling
theory. Though (\ref{eq_27}) is derived for the case of a planet that
occupies a circular orbit and that interacts with test particles on
nearly circular orbits, \citet{qf06} find that the $\mu^{2/7}$ scaling
law holds also for a planet on a moderately eccentric orbit,
interacting with particles on secularly forced eccentric
orbits. \citet{qf06} prefer, however, the coefficient of 1.5 that
originates from the numerical integrations and eccentricity growth
criterion of \citet{dqt89}.\footnote{ \citet{dqt89} also provide a
  simple explanation of behavior within the chaotic zone (for the case
  of circular orbits): particles residing there are perturbed so
  strongly by the planet at each conjunction that successive
  conjunctions occur at uncorrelated longitudes; consequently the
  particle undergoes a random walk in semimajor axis and
  eccentricity, with steps in the walk corresponding to impulses
  at every conjunction.} 
Our work supports the assignment $a_{\rm inner} = a_{\rm chaotic}$
made by Q06.
However, we will calculate still a third
coefficient that best reproduces the HST images
of Fomalhaut's belt; see our equation (\ref{eq_mine}).

The detection of Fom b orbiting just interior to the dust belt
appears
to confirm the expectation of K05, made quantitative by Q06,
that a planet is responsible for truncating the inner edge of the belt
and endowing the belt with its eccentric shape.  However, a closer
examination reveals 
a potential problem with the 
hypothesis that Fom b is solely responsible for the observed belt morphology.
This picture predicts that the planet's orbit be apsidally aligned and coplanar
with (perfectly nested inside) that of the belt. If that were the
case, Fom b's observed position would place it somewhat past orbital
quadrature, near a true anomaly of 109 deg. The problem is that the
expected orbital velocity near quadrature (about 4.2 km/s for
a stellar mass of $2.3 M_{\odot}$ and a semimajor axis of 115 AU)
is lower than K08's estimate
for the actual, deprojected space velocity of Fom b ($5.5_{-0.7}^{+1.1}$ km/s).
The large observed space velocity
suggests that Fom b is
currently located closer to periastron, and that planet and belt
orbits are not apsidally aligned.

Should the apsidal misalignment be taken seriously and the single
planet picture abandoned? It is hard to say.
With Fom b detected at only two epochs so far, it is impossible to
derive a unique orbit. The systematic errors underlying the measured
orbital velocity---in particular frame registration errors that arise
because the star is hidden behind the coronagraphic spot---are
difficult to quantify with confidence. 
More precise statements about Fom b's orbit will have to await future
astrometry.
Our current prejudice is to say that the observational
uncertainties are large enough that apsidal alignment remains a
possibility. Given the observed proximity of Fom b to the belt, it
would be strange if Fom b were not the dominant perturber (see also
the last paragraph of \S\ref{sec_other}).

The open issue of apsidal alignment notwithstanding, we can still place an
upper limit to the mass of Fom b by assuming that it is solely
responsible for shaping the belt. Broadly speaking, more planets
render more of orbital phase space chaotic. Therefore Fom b, if
abetted by other planets, could have a smaller mass than that derived
under the single planet assumption and still truncate
the belt at its observed inner edge. The mass derived
under the single planet assumption would be an upper limit.

Within the single planet picture,
the published dynamical constraint on the
planet's mass is $0.05 \lesssim M_{\rm pl}(M_{\rm J}) \lesssim 0.2$ (Q06).
But there are several reasons to question this result:

\begin{enumerate}
\item Belt properties depend weakly on planet mass, and therefore
  uncertainties in the former are magnified when constraining the
  latter.  For example, $\Delta a_{\rm chaotic}$, which governs the
  relative locations of the belt inner edge and the planet, scales
  only as $M_{\rm pl}^{2/7}$. Similarly weak powers describe how the
  velocity dispersion at the boundary of the chaotic zone depends on
  $M_{\rm pl}$; Q06 parlays this dependence into an upper limit on
  $M_{\rm pl}$, on the grounds that too large a velocity dispersion
  violates the observed sharpness of the belt's inner edge.  We will
  see in our work that this argument does not fully capture the actual
  behavior because it neglects how belt particles located some
  distance from the chaotic zone boundary influence the observed
  sharpness of the edge. In other words, sharpness can only be
  reliably computed with a global model, and not with one that examines the
  chaotic zone boundary exclusively.  See Figure \ref{fig_taunorm} and the
  related discussion in \S\ref{sec_sharp}.
\item 
The published dynamical upper mass limit is based
on the purely gravitational, collisionless behavior of test particles.
But the HST observations of Fomalhaut's belt are of dust grains,
which are influenced by stellar radiation pressure and interparticle
collisions. As explained in \S\ref{sec_oom} 
\citep[see also][]{sc06}, the scattered light observations are likely
dominated by the smallest grains still bound to the star. Of all the solid
material orbiting Fomalhaut, such grains have the largest ratios of
surface area to mass and are the most
seriously affected by radiative forces and collisions.
\item
The lower dynamical mass limit
is also suspect because it is premised on particles colliding
indestructibly and diffusing as members of a viscous circular ring.
The argument underlying the lower limit is that the planet mass cannot be
too small lest dust grains diffuse into the chaotic zone by interparticle
collisions, before the planet can gravitationally eject
them \citep[see also][]{q07}.
But dust particles, colliding at minimum speeds of $\sim$100 m/s (see our
\S\ref{sec_oom}), likely shatter one another. Moreover,
their trajectories are highly elliptical as a consequence
of radiation pressure \citep{sc06}. Their dynamics seem poorly described
by a diffusion equation that conserves particle number,
has constant diffusivity, and assumes circular orbits.
\end{enumerate}

In this work we present a new numerical model of the Fomalhaut dust
belt under the single planet assumption.
It accounts not only for gravitational sculpting by Fom b, but
also for the finite collisional lifetime of dust, the size
distribution of grains, and radiation forces (including
Poynting-Robertson drag, though it is of minor importance compared to
other perturbations).  Though we do not go as far as generating a
model scattered light image to compare with observations, we take the
first step towards this goal by calculating the detailed shape of the
belt's vertical optical depth, $\tau_\perp$, as a function of
semimajor axis $a$.  We compare this optical depth profile with the
corresponding profile of the K05 scattered light model, which in turn was fitted
directly to the HST images.  From this comparison we constrain the
mass and orbit of Fom b: we calculate the possible combinations of
planet mass $M_{\rm pl}$, orbital semimajor axis $a_{\rm pl}$, and
orbital eccentricity $e_{\rm pl}$. 
These results are
independent of any measurement of Fom b, in
particular of its spectrum.  Only as a final extra step do we use information
on Fom b's observed stellocentric distance to see if some of our
mass-orbit combinations might be ruled out.

Our work corroborates the overall picture of Q06---that the planet's
chaotic zone truncates the inner edge of the belt, and that the
planet's orbital eccentricity induces a similar mean eccentricity in
the belt \citep{wyatt99}---but we introduce enough improvements,
especially with regards to our separate handling of unobservable
parent bodies and observable dust grains, that our dynamical
constraint on Fom b's mass supersedes that of Q06.  Moreover,
if future astrometry reveals that that there is no significant
apsidal misalignment, then the precise relationship we derive between
Fom b's mass and its
orbit can be used to determine the former using the latter.
Our model is simple, robust to uncertainties in input parameters, and applicable
to other systems. Though we restrict consideration to a single interior
planet, additional perturbers can be added.

In \S\ref{sec_oom} we present an overview of the Fomalhaut belt,
estimating its physical properties to order-of-magnitude accuracy.
These estimates inform our choices for the input parameters
of our numerical model; that model, and its output, are detailed in 
\S\ref{sec_model}. We summarize and discuss our results---describing
also the curious anomalous acceleration of Fomalhaut
measured by the Hipparcos satellite, and how other planetary companions
in addition to Fom b affect our conclusions---in
\S\ref{sec_sum}.

\section{ORDERS OF MAGNITUDE}\label{sec_oom}

We derive basic properties of the Fomalhaut dust belt, working as much
as possible from first principles and direct observations.
The conclusions reached in the following subsections form the basis
of our numerical model in \S\ref{sec_model}.

\subsection{Bound dust grains present absorbing, geometric cross sections}
Fomalhaut is a spectral type A star of luminosity $L_{\ast} = 16
L_{\odot}$, mass $M_{\ast} \approx 2.3 M_{\odot}$, and age $t_{\rm age} =
200 \pm 100$ Myr (\citealt{byn97}; \citealt{byn98}).  
The circumstellar debris emits a
fractional infrared excess of $L_{\rm IR}/L_{\ast} = 4.6 \times 10^{-5}$
\citep{song01}. K05 report that from 0.6 to 0.8 $\mu$m
wavelengths, the ring has a spatially integrated apparent magnitude of
$m_{\rm app} = 16.2$, nearly entirely due to reflected starlight; for
the star, $m_{{\rm app},\ast} = 1.12$; therefore the fractional reflected
luminosity is $L_{\rm ref}/L_{\ast} \approx 10^{-6}$. If we ignore order
unity effects introduced by anisotropy in the scattering phase
function (see K05 for a fit to this phase function),
the dust albedo is of order $L_{\rm ref}/L_{\rm IR} \approx 0.02$---the
grains are nearly purely absorbing.

Dust is generated by the collisional comminution of larger parent bodies.
The largest parent bodies sit at the top of the collisional cascade
and have lifetimes against collisional disruption equal to $t_{\rm age}$.
We take these largest parents
to occupy a torus of radius $R \approx 140 \AU$,
annular width $\Delta R < R$, and uniform vertical thickness $2H < R$,
to match the observations of K05.
\citet{sc06} refer to this torus as the ``birth ring.''
In principle, a grain of given size that is born into the torus is lost
from it in one of four ways: collisional comminution,
expulsion by radiation pressure, ejection by planetary scattering,
or orbital decay by
Poynting-Robertson (PR) drag.  In practice, for many debris disks, the
first two channels are more important than the fourth
\citep{wyatt05,sc06,tw08}. The ratio of the force of
radiation pressure to that of stellar gravity is
\begin{equation} \label{eq_beta}
\beta = \frac{3L_{\ast}}{16\pi c GM_{\ast} \rho s}
\end{equation}
for a geometrically absorbing grain of internal density $\rho
\approx 1 \gm \cm^{-3}$ and radius $s$, where $c$ is the speed of light
and $G$ is the gravitational constant. 
Grains are unbound from the star
when $\beta \gtrsim 1/2$,\footnote{The critical value of $\beta=1/2$
applies strictly to dust grains released from parent bodies that move initially
on circular orbits. For dust grains released from parent bodies
moving on mildly eccentric orbits, as is the case in Fomalhaut's eccentric belt,
the critical $\beta$ varies with the longitude of release, but is still
near 1/2.} i.e., when \beq s < s_{\rm blow}
\approx \frac{3 L_{\ast}}{8\pi c G M_{\ast}\rho} = 8 \mum \,.
\eeq Because $s_{\rm blow}$ is much greater
than the submicron wavelengths at which the star principally emits, cross
sections for absorption of radiation by bound grains are indeed practically geometric, as
(\ref{eq_beta}) assumes.

\subsection{Radiation Pressure Delivers Grains onto Eccentric, Long-Period Orbits}\label{sec_radpressbirth}

Upon release from parent bodies moving on circular orbits,
bound grains move on orbits of eccentricity \beq e =
\frac{\beta}{1-\beta} = \frac{s_{\rm blow}}{2s - s_{\rm blow}} \label{eq_e}\eeq
and semimajor axis \beq a = \frac{R}{1-e} \,.  \label{eq_a}\eeq
These expressions,
which serve only to guide and which are not used in our more precise
numerical models, ignore the parent-grain relative velocities
with which grains are born, and also any eccentricity
in the parent body orbit. Neither of these errors is serious
for the large grain eccentricities of interest here ($e \gtrsim 1/3$; see \S\ref{sec_tcolnopr}).
The main point is that radiation pressure flings smaller bound grains
born in the torus onto more eccentric, longer period orbits.

\subsection{Collisions Between Grains Are Destructive}\label{sec_destruc}

Colliding belt particles will chip and shatter one another.
For an angular orbital velocity $\Omega_R$ at semi-major axis $R$, the relative
velocity between grains is at least as large as the vertical
velocity dispersion, $\sim H \Omega_R = 100$ m/s. To place
this velocity into some perspective, we note that it is
comparable to the maximum flow speeds of commercial
sandblasting machines.\footnote{See,~e.g.,~http://www.nauticexpo.com/boat-manufacturer/sandblasting-machine-19911.html.}

As a consequence of radiation pressure,
many of the grains will be travelling on bound orbits having eccentricities
on the order of unity (\S\ref{sec_radpressbirth}).
Accounting for orbital eccentricities
(in-plane velocity dispersion) only
increases collision velocities, up to a maximum given by the local
Kepler speed of $R\Omega_R = 4$ km/s. This maximum
is comparable to
elastic wave speeds in rock and will result in catastrophic shattering.

\subsection{Optical Depths: Radial and Vertical}
Since the grains present largely geometric cross sections for absorption
of starlight, $L_{\rm IR}/L_{\ast}$ equals
the fraction of the celestial sphere, centered on the star, that
is subtended by dust grains:
\begin{equation}
\frac{L_{\rm IR}}{L_{\ast}} = \frac{2 \pi R \times 2 H \times \tau_R}{4\pi R^2} = \frac{H}{R}\, \tau_R \,,
\end{equation}
where $\tau_R \ll 1$ is the radial geometric optical depth through the torus.
K05 give a model-dependent aspect ratio of $H/R = 0.025$;
then $\tau_R = 1.8 \times 10^{-3}$.

The vertical optical depth (measured perpendicular to the belt midplane) is
\begin{equation}
\tau_{\perp} = \tau_R \frac{2H}{\Delta R} = \frac{L_{\rm IR}}{L_{\ast}} \frac{2R}{\Delta R} \,,
\end{equation}
independent of $H$. Again from the scattered light observations,
$\Delta R / R \approx 0.17$ (K05), whence
$\tau_{\perp} = 5.4 \times 10^{-4}$.

\subsection{Observable Grains in Fomalhaut's Belt are Bound, and Their Lifetime is Set by Collisions, Not by PR Drag}\label{sec_tcolnopr}
Unbound grains, for which $\beta \gtrsim 1/2$, exit the torus after a
local dynamical time, \beq t_{\rm esc} (s < s_{\rm blow}) \sim
\frac{1}{\Omega_{\rm R}} \sim 200 \yr \,.  \eeq Because the lifetime
of unbound grains, $t_{\rm esc}$, is much shorter than the lifetime of
bound grains---the latter lifetime is set by collisional disruption; see equation
(\ref{eq_tcol}) below---the steady-state population of bound grains will
be proportionately greater than the unbound population. Combining this
fact with the tendency for grain size distributions to concentrate
their collective geometric cross section at the smallest sizes, we
posit that bound grains near the blow-out size, say $s_{\rm blow} < s
\lesssim 2 s_{\rm blow}$, are responsible for much of the observed scattered
light. This view is consistent with the discussion of particle
sizes by K05.

Such grains occupy eccentric orbits, $e > 1/3$,
and are disrupted by collisions amongst themselves.
The lifetime against collisional disruption is $$ t_{\rm col} (s_{\rm blow} < s
\lesssim 2s_{\rm blow}) \sim \frac{1}{\Omega(a) \tau} \sim
\frac{1}{\Omega_R \tau_R} \left( \frac{\Delta R}{R} \right)^{1/2} \nonumber $$
\beq \times \, \frac{1}{(1-e)^{3/2}} \sim 7 \times 10^4 \left( \frac{2/3}{1-e} \right)^{3/2} \yr \,,\label{eq_tcol} \eeq where the effective
optical depth $\tau \sim \tau_R (R/\Delta R)^{1/2}$ accounts for the
path length $\sim$$(R\Delta R)^{1/2}$ traversed by a grain on a highly
elliptical orbit through the birth ring, where densities are highest and
collisions most likely occur. Use of this path length assumes that
relative grain velocities are of order the local Kepler velocity; this
is an acceptable approximation for the order unity eccentricities
of interest here.
Account of the limited fraction of time spent within the torus has
also been made, via $\Omega(a)$ and (\ref{eq_a}).

Compare $t_{\rm col}$ with the Poynting-Robertson drag time, $$
t_{\rm PR}(s_{\rm blow} < s \lesssim 2s_{\rm blow}) = \frac{4\pi c^2
  \rho}{3L_{\ast}}E(e) R^2 s \sim 2.5 \times 10^7 $$ \beq \times \, \left(
  \frac{s}{2s_{\rm blow}} \right) \left( \frac{2/3}{1-e} \right)^{1/2}
\yr \,,\label{eq_tpr} \eeq which is the time for an orbit of initial periastron
$R$ and initial eccentricity $e$ to shrink to a point
\citep{ww50}. This is of the same order as the time for a grain
to have its pericenter be dragged out of the birth ring, for $\Delta R$
not much less than $R$, which is the case here.
The dimensionless function $E(e > 1/3) > 1.9$
quantifies the decay of orbital eccentricity, and diverges as
$(1-e)^{-1/2}$. Comparison of (\ref{eq_tcol}) with (\ref{eq_tpr})
shows that as long as $e$ is not too close to one---i.e., for all
particle sizes outside a tiny interval that just includes $s_{\rm
  blow}$---grains are removed from the ring by collisionally cascading
down to the blow-out size, and PR drag presents only a minor
loss mechanism.  In other words, Fomalhaut's disk is Type B or
collision-dominated \citep{sc06}. 

Our estimate of the collisional
lifetime $t_{\rm col}$ in (\ref{eq_tcol}) informs the duration of our
dust particle simulations, introduced in \S\ref{sec_radrun}.

\subsection{Total Dust Mass Versus Total Parent Body Mass}\label{sec_totparent}
The mass $M_{\rm d}$ in dust responsible for the scattered light is
\beq
M_{\rm d} \sim \frac{8\pi}{3} \rho s \tau_{\perp} R \Delta R \sim 10^{-3} M_{\oplus} \left( \frac{s}{2s_{\rm blow}} \right)\,.
\eeq
The mass $M_{\rm pb}$ in the largest parent bodies at the top of the collisional
cascade is given by the steady-state condition
\beq
\frac{M_{\rm pb}}{t_{\rm age}} \approx \frac{M_{\rm d}}{t_{\rm col}}
\eeq
which implies
$$M_{\rm pb} \sim 3 M_{\oplus}\,.$$
This is a minimum mass for the disk as a whole
because still larger bodies may exist
which are collisionless over $t_{\rm age}$.

Some workers \citep[e.g.,][]{backman93} calculate
the mass in parent bodies by explicitly assuming a size distribution
appropriate for an idealized collisional cascade \citep{dohn69} and taking the upper size to be some value $> 1$ km. Not only is it unnecessary to specify
a size distribution, but it is not justified to adopt a specific value
for the parent body size without first establishing that a typical
such body can be collisionally disrupted within the finite age
of the system. The super-kilometer sizes often invoked (e.g., K05) fail this test.

\section{NUMERICAL MODEL}\label{sec_model}

We devise a dynamical model of the Fomalhaut planet-belt system
that reproduces approximately some of the properties inferred
from the HST observations. We compute the shape of the vertical
optical depth profile, $\tau_\perp (a)$, of dust particles in the belt
and match this profile against that of the K05 scattered light model.
We seek in particular to find those combinations of planet mass and orbit
that yield an inner edge to the belt of $a_{\rm inner} = 133 \AU$.

As stressed throughout this paper, the model assumes
only a single interior planet apsidally aligned with the belt,
an idealization not supported
by Fom b's nominal space velocity (\S\ref{sec_intro}) or by
the Hipparcos data (\S\ref{sec_other}). Nevertheless,
the planet masses derived from the single planet model
can usefully be interpreted as upper limits (\S\ref{sec_intro}).

The procedure is detailed in \S\ref{sec_procedure};
results are given in \S\ref{sec_result}; and
extensions of the model, including some validation tests,
are described in \S\ref{sec_extend}.

\subsection{Procedure}\label{sec_procedure}

Our numerical modeling procedure divides into four steps,
described in the following four subsections, \S\S\ref{sec_parent}--\ref{sec_compare}.
In short, we (1) create a population of parent bodies
that is stable to gravitational perturbations from Fom b over $t_{\rm age}$;
(2) release dust particles from stable parent bodies
and follow dust trajectories in the presence of radiation forces
over the collisional lifetime $t_{\rm col}$; (3) compute the optical
depth profile $\tau_\perp (a)$ of dust particles at the end
of $t_{\rm col}$, accounting for their size distribution; and (4) compare with
the K05 scattered light model, which serves as our proxy for the observations.

\subsubsection{Step 1: Create Stable Parent Bodies}\label{sec_parent}
Parent bodies (a) execute orbits that are stable to gravitational
perturbations over $t_{\rm age}$, (b) occupy the top of the
collisional cascade, which by definition implies that their orbits are
little affected by catastrophic collisions for times $t < t_{\rm age}$,
and (c) are large enough that radiation forces are negligible.
We assume further that (d) the self-gravity of the belt
is negligible.  Thus
the problem of simulating a realistic parent body is a purely
gravitational, three-body problem involving the star, planet, and
exterior parent body, where the parent body is treated as a test
particle.
Finding stable test particle orbits is a straightforward
matter of specifying their initial conditions, integrating the equations of
motion forward for $\sim$$t_{\rm age}$, and selecting those particles
that survive the integration.

Integrations of parent bodies are performed
with the \texttt{swift\_whm} code, written by H.~Levison and M.~Duncan
(www.boulder.swri.edu/\~{}hal/swift.html) using the
\citet{wh91} algorithm.
We set the stellar mass $M_{\ast} = 2.3 M_{\odot}$. 
In each of our five mass models,
a planet of mass $M_{\rm pl} \in (0.1, 0.3, 1, 3, 10) M_{\rm J}$
resides on a (fixed) elliptical orbit
of semimajor axis $a_{\rm pl}$ and eccentricity
$e_{\rm pl}$. These quantities are listed in
Table 1 (placed at the end of this manuscript); see below for how we relate $e_{\rm pl}$ to $a_{\rm pl}$,
and Step 4 (\S\ref{sec_compare}) for how we select $a_{\rm pl}$ given
$M_{\rm pl}$.
The planet's orbital plane defines the $x$-$y$
reference plane for the system. The planet's longitude of periastron
$\pomega_{\rm pl} = 0$. At the start of the integration,
the planet is located at periastron.
All orbital elements reported here are stellocentric and osculating unless
otherwise noted.

Each run begins with $N_{\rm tp} = 10^4$ test particles and lasts $10^8$ yr.
Particles are discarded as ``unstable'' if either they approach within
a Hill sphere $R_{\rm H} = (\mu/3)^{1/3}a_{\rm pl}$
of the planet, or their distance from the star exceeds 1000 AU, as a
result of gravitational scatterings off the planet. Particles that survive
until the end of the run are deemed ``stable'' and are used
in subsequent steps of our procedure.
The integration timestep is 15000 days, equivalent to 5--7\% of the planet's
orbital period, depending on the model.

Initial conditions for test particles are as follows.  Initial
semimajor axes $a$ are distributed uniformly
between $120\AU$ and $140 \AU$ (for the 0.1 $M_{\rm J}$ model,
the starting particle semimajor axis is 125 AU since the planet semimajor
axis turns out to be $a_{\rm pl} = 120\AU$).
Initial eccentricities are set equal
to the forced values given by the classical, linear, secular
theory of Laplace-Lagrange (L-L):

\beq e(a) = e_{\rm forced}(a) =
\frac{b^{(2)}_{3/2}(a_{\rm pl}/a)}{b^{(1)}_{3/2}(a_{\rm pl}/a)}
e_{\rm pl} \label{eq_ll} \eeq
where the $b$'s are the usual Laplace coefficients
\citep[e.g.,][]{md00}.  Initial inclinations $i$ of test particles
are drawn randomly from a uniform distribution between 0 and 0.025
rad.\footnote{Such an inclination distribution is unphysical because
in reality, there is zero probability density for finding an object
with zero inclination.  Nevertheless, we adopt our boxcar distribution
for simplicity.}  Our maximum inclination matches the opening angle of
the K05 scattered light disk model. Initial longitudes of
periastron of all particles equal the secularly forced value
$\pomega = 0$, corresponding to apsidal alignment with the
planet; longitudes of ascending node $\widetilde{\Omega}$ are
drawn randomly from a uniform distribution between 0 and $2\pi$; and
arguments of periastron $\omega = -\widetilde{\Omega}$ (so that
$\pomega = 0$). Finally, initial mean anomalies $M$ are drawn randomly
from a uniform distribution between 0 and $2\pi$.

For a given $a_{\rm pl}$ in a given model, the planet's eccentricity $e_{\rm
pl}$ is such that a test particle at $a=140.7\AU $ acquires a secularly
forced eccentricity of $e = 0.11$, as computed using (\ref{eq_ll}). Such parameters are inspired by the elliptical orbit fitted by K05 to 
the brightest portions of the belt.
The planetary eccentricity so chosen is about $e_{\rm pl} = 0.12$,
varying by up to 15\% between our five models (see Table 1).

According to L-L, the initial conditions so prescribed produce test
particle orbits whose eccentricity vectors $\mathbf{e} = (e\cos
\pomega, e\sin \pomega)$ are purely forced; they have and will have no free
component \citep[e.g.,][]{md00}.  As such, because the planet's orbit
is fixed, belt particle orbits also should not vary, at the L-L
level of approximation.\footnote{Laplace-Lagrange
truncates the secular disturbing
function at $O(e^2,i^2)$ and so in reality and in numerical integrations, the
eccentricities and apsidal angles are still expected to vary somewhat with
time with our initial conditions, even if the system were purely secular.
One can avoid such truncation error by employing
Gauss's perturbation equations for $\dot{e}$ and $\dot{\pomega}$
and integrating the
planetary forces over Gaussian wires \citep[e.g.,][]{md00},
thereby finding exact
forced eccentricities for which $\dot{e}=\dot{\pomega}=0$. We skip
this refinement, in part because the system is not purely secular;
nearby mean-motion resonances influence the dynamics.}
In \S\ref{sec_parentdyn}, we describe the
extent to which this expectation is borne out.

Our initial conditions, which comprise
nested, apsidally aligned, purely forced elliptical orbits,
are designed to reproduce the observed elliptical belt of Fomalhaut
\citep{wyatt99,quillen06}.
However, such forced orbits
are not the only ones that are stable
in the vicinity of Fom b. In \S\ref{sec_resonance}, we experiment
with a different set of initial conditions that generate
another class of stable parent body.

\subsubsection{Step 2: Integrate Dust Trajectories}\label{sec_radrun}
Having created an ensemble of stable parent bodies,
we now model the dust generated by such bodies.
At the end of a $10^8$-yr-long integration from Step 1, 
we take each stable 
parent body and have it ``release'' a dust grain with
the same instantaneous position
and velocity as its parent's. Each dust grain's trajectory
is then integrated forward under the effects of
radiation pressure and PR drag. That is,
in addition to the gravitational accelerations
from the star and the planet, a dust particle also
feels a radiative acceleration \citep[see, e.g.,][]{burns79}
\beq
\mathbf{a}_{\rm rad} = \frac{GM_{\ast}\beta}{r^2} \left( \hat{\mathbf{r}} - \frac{v_r \hat{\mathbf{r}} + \mathbf{v}}{c} \right)
\eeq
where 
$\mathbf{r} = r \hat{\mathbf{r}}$ is the vector displacement from the
star to the grain, $\mathbf{v}$ is the velocity of the grain relative
to the star, and $v_r = \mathbf{v} \cdot \hat{\mathbf{r}}$ accounts for
the Doppler shift in stellar radiation seen by the grain.
We add this radiative acceleration
to the Bulirsch-Stoer (B-S) integrator \texttt{swift\_bs}, written by
Levison \& Duncan. We prefer to modify the B-S integrator
over the Wisdom-Holman integrator, since the latter
is designed to model a dissipationless Hamiltonian system; when
PR drag is included, the system is dissipative, and it is not obvious
to us how we should add the radiative perturbations
to the symplectic kick-drift-kick algorithm. (Nevertheless, adding radiative
forces to symplectic integrators is standard practice
in the literature and has been shown to produce accurate results.)
Though the B-S integrator is slower, it is fast enough for our purposes
since our integration times for Step 2 are short (see below).
The accuracy parameter ``eps'' of \texttt{swift\_bs} is set
to $10^{-8}$. The modified code was tested on test
particles having various $\beta$'s, producing results for radiation blow-out
and PR drag in excellent agreement with analytic and
semi-analytic studies \citep[e.g.,][]{ww50}.

From each of the five parent body integrations completed in Step 1,
we generate
eight dust simulations in Step 2, each characterized by a single value of
$\beta \in (0, 0.00625, 0.0125, 0.025, \ldots, 0.4)$.
Dust grains released with such $\beta$'s are mostly
still bound to the star, albeit on highly eccentric orbits
for $\beta$ approaching 0.4. Bound grains contribute
substantially, if not predominantly, to the scattered light
observations: as sketched in \S\ref{sec_oom}, because $t_{\rm PR} \gg t_{\rm col}
\gg t_{\rm esc}$, the lifetime of bound grains
is set by destructive interparticle collisions, not
by PR drag, and the steady-state population of bound grains greatly
outweighs that of unbound grains, by $t_{\rm col}/t_{\rm esc}$.

Because the dust lifetime is set by collisions,
we extract dust grain orbits for further analysis in Step 3
after integrating for a time $t_{\rm col}$.
Following our order-of-magnitude estimate
(\ref{eq_tcol}),
we set $t_{\rm col} = 10^5 \yr$. During the integration,
we discard particles that approach
within a Hill sphere of the planet or whose distances from the star
exceed 10000 AU (this is a factor of 10 larger than the cut imposed
in Step 1, because large apastron distances result from the onset of radiation
pressure and do not necessarily imply orbital instability).
As Table 1 indicates, few if any dust particles are discarded in
our $\beta$-simulations, with the exception of our 10 $M_{\rm J}$ model.

In \S\ref{sec_tcol},
we test the sensitivity of our results to $t_{\rm col}$,
whose value we know only to within factors of a few.
In that subsection we also examine whether our results change
significantly if we model
the dust more realistically by releasing grains
gradually over $t_{\rm col}$.

\begin{figure}
\centering
\scalebox{0.8}{\plotone{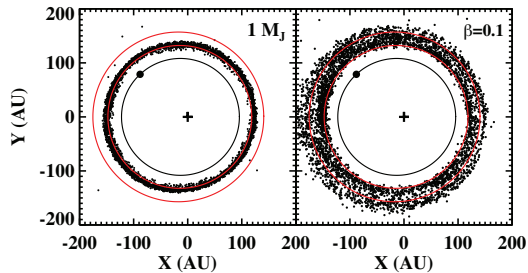}}
\caption{Snapshots of parent bodies (left) and $\beta=0.1$ dust grains (right),
for our $1 M_{\rm J}$ model. The cross marks Fomalhaut, while the
solid circle marks Fomalhaut b. Parent bodies are imaged after an
integration time of $t_{\rm age} + t_{\rm col}$. Dust particles are released
from parent bodies with zero relative velocity after $t_{\rm age}$
and their trajectories integrated forward with $\beta = 0.1$ for $t_{\rm col}$.
Red ellipses correspond to the inner and outer boundaries of the K05 scattered
light model ($a_{\rm inner} = 133 \AU$, $a_{\rm outer} = 158\AU$, $e = 0.11$).
Radiation pressure spreads dust particles outward from where they were born,
but leaves their inner boundary practically coincident with that of parent
bodies.
}
\label{fig_snap100}
\end{figure}

Figure \ref{fig_snap100}
provides sample snapshots of dust grains and their parent bodies
projected onto the $x$-$y$ plane, for our $1 M_{\rm J}$ model.

\subsubsection{Step 3: Compute Optical Depth Profile}\label{sec_tau}

If we had a sufficiently large number of dust particles, we could simply
take a snapshot of each Step 2 $\beta$-simulation after $t_{\rm col}$ and count the
number of dust particles per unit $x$-$y$ area of the belt. We would
thereby measure
the surface density $N_{\beta} (x,y)$, and by extension the vertical optical
depth $\tau_{\perp} (x,y)$. In practice, we do not have
enough particles, and such an exercise produces noisy results.

We greatly improve the signal-to-noise by
spreading each dust particle along its orbit according
to how much time it spends traversing a given segment of its orbit.
In other words, we replace each dust particle with its
equivalent Gaussian wire, and measure the optical depth
presented by the collection of wires. We refer the Kepler
elements of a dust particle's orbit to $(1-\beta)$ times
the stellar mass; otherwise the elements would not remain constant in a two-body
approximation.

First we construct an eccentric grid by partitioning the $x$-$y$ plane into a
series of nested, confocal ellipses, all having the same eccentricity of $0.11$
(K05), and having semimajor axes running uniformly from $a_1=100\AU$
to $a_N = 220\AU$ in steps of $\Delta a = 0.5 \AU$.  
Our goal is to measure $\tau_\perp (a_i)$,
the average optical depth between the $i$$^{\rm th}$ ellipse having
semimajor axis $a_i$ and the $(i+1)$$^{\rm th}$ ellipse having
semimajor axis $a_{i+1} = a_i + \Delta a$.  Each dust particle orbit
is divided into 1000 segments spaced uniformly in true anomaly from 0
to $2\pi$. Each segment maps to a certain location on the grid (i.e.,
the $x$-$y$ position of each segment falls between two adjacent
ellipses on the grid). Associated with each segment is an orbital
weight, equal to the fraction of the orbital period spent traversing
that segment (the sum of all orbital weights for a given
particle/orbit equals one). The orbital weight for each
segment is added to its grid location. This process is
repeated over all segments of all orbits. Finally, at each grid
location $a_i$, the sum of
orbital weights is divided by the area of the
annulus extending from the $i$$^{\rm th}$ ellipse to the
$(i+1)$$^{\rm th}$ ellipse. This yields the relative surface density
profile $N_{\beta} (a_i)$, for a simulation characterized by
$\beta$.

\begin{figure}
\centering
\scalebox{0.7}{\plotone{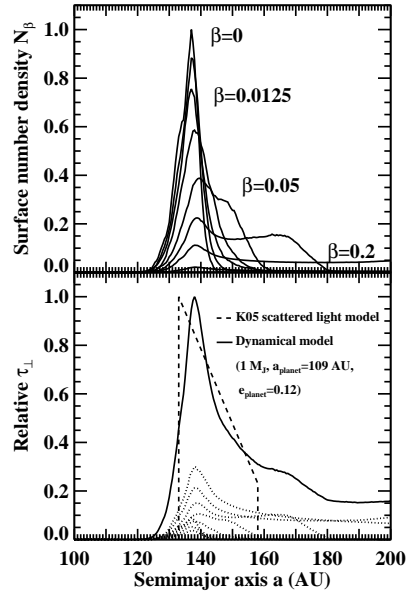}} 
\caption{{\it Top panel}: Surface density profiles of dust grains
for our $1 M_{\rm J}$ model,
computed $t_{\rm col} = 10^5\yr$ after release, normalized
to the peak of the $\beta=0$ curve. These profiles
are computed by binning wire segments on a fixed elliptical grid,
as described in the main text under Step 3 of our procedure.
Profiles shrink vertically and widen horizontally with increasing $\beta$,
reflecting how increased radiation pressure spreads particles outward
by amplifying apastron
distances. By contrast, periastron distances are more nearly
conserved for bound particles, since they always return to the birth ring
regardless of the strength of radiation pressure. Thus the peaks
of the $N_\beta$ profiles, which mark the location of the birth
ring of parent bodies (see left panel of Figure \ref{fig_snap100}),
hardly shift with $\beta$.
{\it Bottom panel}: Vertical optical depth profile (solid line)
obtained by adding together
the individual $N_\beta$ profiles (dotted lines),
appropriately weighted according to a
Dohnanyi size distribution (see equation \ref{eq_linearcombo}).
At $a \lesssim 160 \AU$, the profile resembles
that of the K05 scattered light model (dashed line),
which itself is an approximate, idealized, and non-unique representation of
the HST
observations. Discrepancies at large $a\gtrsim 160 \AU$ are expected,
in large part because the HST images are too noisy to usefully constrain
the K05 model there.}
\label{fig_tau_combine}
\end{figure}

The various $N_{\beta}$ profiles
for our $1 M_{\rm J}$ model are plotted in the top panel of
Figure \ref{fig_tau_combine}. The profiles are
normalized to the peak of the $N_0$ ($\beta=0$) profile.
Since the number of dust particles 
is practically constant across all $\beta$-simulations
within a given mass model (Table 1),
the decreasing height of each
$N_{\beta}$ profile with increasing $\beta$ simply reflects
how dust particle orbits become increasingly eccentric and elongated with
increasing $\beta$ (cf. equation \ref{eq_e}).
In other words, the same number of particles
is being spread into a disk that extends farther out the greater the radiation
pressure. At the same time, the peaks of the $N_{\beta}$
profiles hardly move with increasing $\beta$: as long as the grain is still
bound to the star, it must always return to the same stellocentric
distance at which it was
released, no matter how strongly it feels radiation pressure.
That release distance is located in the birth ring \citep{sc06}
or, more accurately, the birth ellipse of parent bodies,
near $a\approx 130$--$140\AU$;
see the left-hand panel of Figure \ref{fig_snap100}.

Once all the $N_{\beta}$ profiles are in hand, we combine them
linearly to produce the total optical depth profile $\tau_{\perp}$:

$$ 
\tau_{\perp} = \sum_{\beta\neq 0} N_{\beta} \frac{\max N_{0.00625}}{\max N_{\beta}} \left( \frac{\beta}{0.00625} \right)^{q-3} $$ \beq + \, N_{0} \frac{\max N_{0.00625}}{\max N_0}(1+\sqrt{2})\,.
\label{eq_linearcombo} \eeq

\noindent The rationale behind this formula is as follows.  As Figure
\ref{fig_tau_combine} indicates, the maxima of the $N_{\beta}$ profiles
are all situated in the birth ring. Following \citet{sc06}, we posit
that the size distribution of grains in the birth ring is given by a
\citet{dohn69} cascade, with differential power-law index $q = 7/2$. For
such a power-law size distribution, the collective surface area or
geometric optical depth, evaluated per logarithmic bin in $\beta$,
scales as $\beta^{q-3}$ ($\propto s^{3-q}$).  The two factors
multiplying $N_{\beta}$ in the sum in (\ref{eq_linearcombo}) enforce
this scaling in the birth ring (whose location is traced by the
maxima of the surface density profiles), and we have adopted the
$\beta=0.00625$ bin as our reference bin.

The last term proportional to $N_0$ accounts for the optical depth
contributed by grains having $0 < \beta < 0.00625$. We take the
surface density profile of such grains to be given by $N_0$; this is a
good approximation, as there is little difference between
$N_{0.00625}$ and $N_0$ (Figure \ref{fig_tau_combine}, top panel).
Since our grid of models for $\beta \geq 0.00625$ is logarithmic in
$\beta$---successive bins are separated by factors of $B=2$---we
extend the same logarithmic grid for $\beta < 0.00625$.  Then the
optical depth coming from all grains having $\beta < 0.00625$, scaled
to the optical depth in our $\beta = 0.00625$ reference bin, is
$\sum_{j=1}^{\infty} (1/B)^{j(q-3)} = 1 + \sqrt{2}$.

\subsubsection{Step 4: Compare with Observations}\label{sec_compare}
A rigorous comparison with observations would require us to produce a
scattered light image based on our dynamical model. This is a
considerable task. Our $\tau_\perp$ profile, combined with the
vertical density distribution and a grain scattering phase function,
would be used to calculate the direction-dependent emissivity of the
belt as a function of 3D position. This emissivity model would then be
ray-traced at a non-zero viewing angle to produce a model scattered
light image. Various parameters (e.g., normalization of $\tau_\perp$,
grain scattering asymmetry parameter, viewing angle) would need to be
adjusted, including input parameters from Steps 1--3 (distribution of
initial semimajor axes, distribution of initial inclinations), to
produce a good subtraction of the observed image.

In this first study, we attempt none of this. Instead we compare the
$\tau_\perp$ profile given by (\ref{eq_linearcombo}) with the
corresponding optical depth profile of the K05 scattered light model,
focussing on the one belt property that seems most diagnostic of
planet mass and orbit: the belt's inner edge. The K05 optical depth
profile extends from $a_{\rm inner} = 133\AU$ to $a_{\rm outer} =
158\AU$ and falls as $a^{-8.5}$ (their fitted volumetric
number density of grains
falls as $a^{-9}$, while the fitted vertical thickness of the belt
increases as $a^{0.5}$).  In our dynamical modeling, for given planet
mass $M_{\rm pl}$, we adjust only the planet semimajor axis $a_{\rm
  pl}$ and repeat Steps 1--3 until the minimum semimajor axis at which
$\tau_\perp$ reaches half its maximum value---the ``half-maximum
radius''---equals $a_{\rm inner} = 133 \AU$.

Since our dynamical models are anchored to $a_{\rm inner}$, we should
have a sense of the uncertainty in this parameter.  The K05 model is
based on fits ``by eye.'' From the visual fits, the uncertainty is
about $\pm 1$ AU, slightly larger than the size of a 2-pixel
resolution element (0.1 arcsecond or 0.77 AU; the images are binned $2
\times 2$ before they are modeled, to increase signal-to-noise).
The error in $a_{\rm inner}$ propagates into our constraints
on planet mass.

It is well to appreciate that while our general goal is to
reproduce the shape of the optical depth profile of the K05 scattered
light model, that model is itself highly idealized, characterized
by knife-edge sharp inner and outer edges and simple power-law behavior.
Fitting by eye means the model is at best approximate.
And as K05 caution, only a restricted azimuth of the belt (near their
quadrant ``Q2'') was analyzed in detail to produce their fit
parameters.  Therefore we should neither aim for, nor expect, perfect
agreement between our dynamical model and the K05 model.  Our task
instead is to use the K05 model as a guide, to identify robust trends
and to rule out those regimes of parameter space for Fom b that give
blatantly poor agreement with observed belt properties.

Lastly, neither our dynamical model nor the K05 scattered light model
should be trusted at large $a\gtrsim 160 \AU$. At large distance,
barely bound grains whose $\beta$'s are arbitrarily close to the
blow-out limit of $\sim$0.5 dominate $\tau_\perp$. Our set of 8
$\beta$-simulations lacks the resolution in $\beta$ to accurately
model this outer disk (see \citealt{sc06} for an analytic treatment
appropriate for a circular birth ring).  Observationally, the HST
images at stellocentric distances $\gtrsim 158 \AU$ are dominated by
the sky background; see Figure 3 of K05.

\subsection{Results}\label{sec_result}

In \S\ref{sec_newcoeff} we give an approximate formula relating the
planet mass to the width of the chaotic zone, based on our five mass
models.  In \S\ref{sec_res_tau}, we compare our optical depth profiles
with that of the K05 scattered light model.  Based on this comparison
we argue against large planetary masses for Fom b. In \S\ref{sec_apo}
we argue the same point by comparing our model planetary orbits with
the observed stellocentric distance of Fom b. Finally, the underlying
dynamics of parent bodies and of dust particles is discussed briefly
in \S\ref{sec_parentdyn}.

\subsubsection{Chaotic Zone Width}\label{sec_newcoeff}

\begin{figure}
\centering
\scalebox{0.8}{\plotone{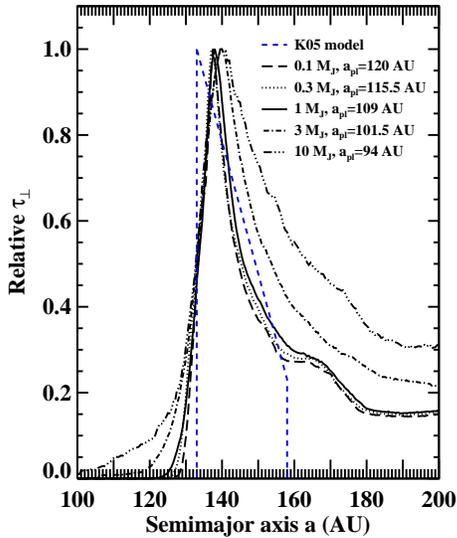}} 
\caption{Vertical optical depth profiles of our five mass models (black lines),
overlaid with that of the K05 scattered light model (blue dashed line).
Parameters for our dynamical models are listed in the inset and are such
that the ``half-maximum radius''---the minimum semimajor axis for which
$\tau_\perp$ attains half its maximum value---equals 133 AU, the innermost
semimajor axis of the K05 model. Models for which $M_{\rm pl} \leq 1 M_{\rm J}$
do equally well in reproducing the K05 model. As $M_{\rm pl}$ increases,
the $\tau_\perp$ profiles widen because the planet increasingly perturbs
dust grains onto eccentric orbits. The $10 M_{\rm J}$ model is probably
unacceptably wide.
At $a\gtrsim 160 \AU$, neither the dynamical model nor the K05 model
is trustworthy; the former suffers from poor resolution in $\beta$,
while the latter is limited by sky background (see Figure 3 of K05).
}
\label{fig_tau_triple}
\end{figure}

In Figure \ref{fig_tau_triple}, we overlay the $\tau_\perp$ profiles
of our five mass models together with the optical depth profile of
the K05 scattered light model. As described in \S\ref{sec_compare},
the planet semimajor axis $a_{\rm pl}$ for each mass model is chosen
such that the ``half-maximum radius''---the smallest semimajor axis at which
$\tau_\perp$ attains half its peak value---equals K05's
$a_{\rm inner} = 133 \AU$.  The $a_{\rm pl}$'s so determined are
listed in Table 1 
and also annotated on Figure
\ref{fig_tau_triple}. They are such that the semimajor axis separation
between belt inner edge and planet is given by
\beq a_{\rm inner} - a_{\rm pl} = 2.0\,\mu^{2/7} a_{\rm pl} \,,\label{eq_mine}
\eeq
with less than 6\% variation in the coefficient across
mass models, and where $a_{\rm inner} = 133 \AU$.
Because we are connecting more directly to the HST observations,
our Fomalhaut-specific
coefficient of 2.0 is preferred over the smaller coefficients
cited by previous works; accuracy matters for determining planet mass,
whose value scales strongly as the 7/2 power of distance measurements.
Measured in Hill
radii (evaluated using $a_{\rm pl}$),
the separation $a_{\rm inner} - a_{\rm pl}$ ranges from 3.7
  to 4.5 $R_{\rm H}$ in order of decreasing $M_{\rm pl}$.

\subsubsection{Comparison of $\tau_\perp$ Profiles Implies $M_{\rm pl} < 3 M_{\rm J}$}\label{sec_res_tau}

What resemblance there is in Figure \ref{fig_tau_triple} between
dynamical and scattered light $\tau_\perp$ profiles leads us to believe
that we are on the right track towards understanding the underlying properties
of the Fomalhaut planet-belt system.
We are especially encouraged when we consider that with the 
exception of $a_{\rm pl}$, none of our model parameters (e.g., grain
size index $q$, distribution of initial semimajor axes) has been
adjusted from its naive standard value. And as we emphasized
at the end of \S\ref{sec_compare}, the K05 scattered light model
is itself highly idealized and approximate, and does not represent
a unique model of the observations. In particular, the K05 model
idealizes the inner edge as a step function, but smoother profiles
also seem possible; the degree of smoothness may help to constrain $q$.

Discrepancies at
large $a \gtrsim 160 \AU$ are not serious, since both the dynamical
and scattered light models are known to fail there (\S\ref{sec_compare}).
The deficiency in our dynamical model can be remedied
by having a finer grid in $\beta$ near the blow-out value of $\sim$0.5,
while improvements in the scattered light model await deeper imaging campaigns.

\begin{figure}
\centering
\scalebox{0.8}{\plotone{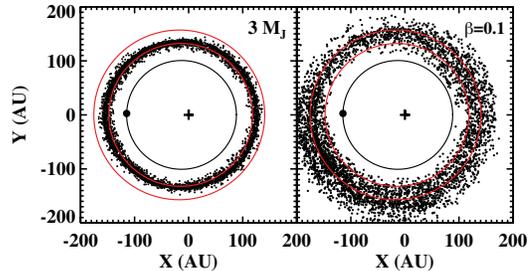}}
\caption{Snapshots of parent bodies (left) and $\beta=0.1$ dust grains (right),
for our $3 M_{\rm J}$ model. Compare with the $1 M_{\rm J}$ model shown in Figure
\ref{fig_snap100}. The larger the planet mass, the more
dust particles are perturbed by the planet into a more spatially extended
disk.
}
\label{fig_snap300}
\end{figure}

As $M_{\rm pl}$ increases, the $\tau_\perp$ profiles broaden---see in
particular the curve for $10 M_{\rm J}$.  Upon release, dust particles
find themselves in a weakened stellar potential because of radiation
pressure. More massive planets more readily perturb
dust onto more eccentric orbits that extend both inside and outside of
the birth ring.  This is further illustrated in Figure \ref{fig_snap300},
which shows a snapshot of $\beta = 0.1$ grains for $M_{\rm pl} = 3 M_{\rm
  J}$.  Moreover, as documented in Table 1,
planetary perturbations are so severe for our $10 M_{\rm J}$
model that about 1/3 of the dust particles launched with $\beta=0.05$ are
discarded as unstable within $t_{\rm col} = 10^5\yr$. 
(Why $\beta=0.05$? Larger $\beta$ grains are, upon
release, repelled immediately away from the planet onto highly
eccentric trajectories by radiation pressure alone
and are therefore less likely to be rendered unstable by the planet.
Smaller $\beta$ grains share essentially the same stability
properties as the parent bodies.)

The $\tau_\perp$ profile for $M_{\rm pl} = 10 M_{\rm J}$ is probably
unacceptably broad: at $140 \AU \lesssim a \lesssim 160 \AU$, the
dynamical model predicts too large an optical depth compared to the
K05 model, by about a factor of two.  At these distances,
a factor of two overluminosity in the belt
is not easily accommodated, as judged from the error bars on
the observed surface brightness profile---see Figure 3 of K05.
This same $10 M_{\rm J}$ model
also produces a tail extending inward to $a\lesssim 120 \AU$, but the
observations, whose dynamic range in surface brightness is limited to
less than 10:1 (see Figure 3 of K05), probably cannot rule out such a
feature. We have verified that the excessively large $\tau_\perp$ at
$a \gtrsim 140\AU$ follows primarily from the large eccentricities
acquired by $\beta\approx 0.2$--0.4 dust particles from planetary
perturbations.
By contrast, the
$M_{\rm pl} \leq 1 M_{\rm J}$ models, which produce
practically identical $\tau_\perp$ profiles, appear compatible
with the K05 model, given the various limitations of the latter.

The low-mass models having $M_{\rm pl} \leq 1 M_{\rm J}$ have 
inner edges that are practically identically sharp. A measure
of the sharpness is the distance over which $\tau_\perp$
rises from the half-maximum radius to the semimajor axis
at which $\tau_\perp$ peaks,
normalized by the half-maximum radius. This fractional distance is
$\delta = 4.5\AU/133\AU = 0.034$. \citet{quillen06}
recognizes that in fitting the observed drop-off
in surface brightness for a belt that is
inclined to our line of sight, 
a trade-off exists between the belt's vertical and radial density
profiles. Either the belt can have a finite vertical thickness and be knife-edge
sharp in the radial direction (as in the K05 scattered light model);
or have zero vertical thickness and drop off gradually in the radial direction;
or be characterized by some intermediate combination.
In this context, our measure for the fractional radial drop-off distance,
$\delta = 0.034$, compares promisingly with the fractional vertical drop-off
distance inferred by K05, $2H/R = 0.025$. In the future, we will have
to adjust both radial and vertical drop-off distances to better
reproduce the scattered
light observations. The vertical thickness of the belt appears to be
fairly easily adjusted in our model, given that the inclinations
of our dust particles are mostly unchanged from the assumed
initial inclinations of our parent bodies (see Figure \ref{fig_inc} 
and related discussion in \S\ref{sec_parentdyn}).

In the bottom panel of Figure \ref{fig_tau_combine}, we plot as dotted
curves the separate terms in equation (\ref{eq_linearcombo}) that add
up to $\tau_\perp$.  The terms corresponding to larger $\beta$ (or
smaller grain size $s \propto 1/\beta$) dominate, as a consequence of
our assumption that the grain size distribution in the birth ring
follows a Dohnanyi law. That law apportions the bulk of the geometric
surface area in the smallest grains.

\subsubsection{Fom b's Current Stellocentric Distance Implies $M_{\rm pl} < 3 M_{\rm J}$}\label{sec_apo}

Each of our mass models specifies a certain orbit for Fom b (Table 1) that is
tuned, through multiple iterations of Steps 1--4, to
generate the observed ellipticity of the belt and to yield a half-maximum
radius equal to K05's $a_{\rm inner} = 133 \AU$. If the apocentric
distance of a model orbit is less than the observed stellocentric distance
of Fom b, then that model can be ruled out.

Currently, only the distance between Fom b and its host star projected
onto the sky plane is known. In 2006, that projected distance was 97.6
AU.  Inferring the true stellocentric distance requires that we
de-project the orbit of Fom b. But with only two epochs of
observation, a unique de-projection is not possible.  Nevertheless, we
can perform a rough de-projection by making a few assumptions. We
assume that the planet's orbit is oriented such that the line of nodes
between the planet's orbit and the sky plane coincides with the line
joining the observed belt ansae.  We also assume that the inclination
of the planet's orbit to the sky plane equals $65.6^{\circ}$, the same
as that inferred for the belt orbital plane (K05).  These assumptions
do not restrict the apsidal orientation of the planet's orbit.

The resultant de-projected stellocentric distance of Fom b in 2006 is
119 AU, with a systematic error of probably no more than a couple
AUs. Such a distance argues against our $10 M_{\rm J}$ model, for
which the apocentric distance of Fom b is 107 AU. The discrepancy is
depicted in Figure \ref{fig_parent}.  That same figure shows that the
$3 M_{\rm J}$ model is also inconsistent, but only marginally so, and
without a proper de-projection, we hesitate to rule it out.  Lower
masses $M_{\rm pl} \leq 1 M_{\rm J}$ are, by contrast, easily
compatible.  These findings reinforce those based purely on a
comparison of the optical depth profiles (\S\ref{sec_res_tau}).

\subsubsection{Parent Body and Dust Particle Dynamics}\label{sec_parentdyn}

\begin{figure}
\centering
\scalebox{0.8}{\plotone{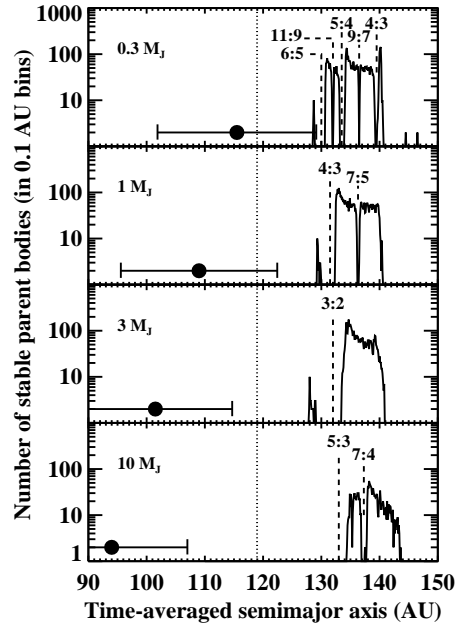}} 
\caption{Histogram of time-averaged semimajor axes of parent bodies
that survive for $10^8\yr$ in the vicinity of Fom b.
The bin width is 0.1 AU, and the time average is performed over $10^5 \yr$.
Each panel corresponds to a different mass model, as annotated.
The black circle in each panel marks the semimajor axis of Fom b,
with error bars extending from the model orbit's pericentric distance
to its apocentric distance. Only model orbits corresponding to
$M_{\rm pl} \lesssim 1 M_{\rm J}$ are consistent with the estimated
de-projected stellocentric distance of Fom b in 2006 (dotted vertical line).
Stable parent bodies are located outside the planet's chaotic zone, at semimajor
axes greater than the planet's by about $2\mu^{2/7}a_{\rm pl}$. Inside the
chaotic zone, first-order mean-motion resonances overlap and particle orbits
are short-lived. Outside
the chaotic zone, parent bodies reside stably on secularly forced eccentric
orbits, with occasional gaps located at mean-motion resonances as indicated.
The gaps are evacuated for reasons likely analogous to why the solar
system's Kirkwood gaps are empty of asteroids.
}
\label{fig_parent}
\end{figure}

In Figure \ref{fig_parent}, we supply sample histograms of time-averaged
semimajor axes of stable parent bodies generated in Step 1.
The time average is performed over a $10^5$-yr-long window (with $\beta=0$)
following each $10^8$-yr-long integration. Intriguingly,
parent bodies appear evacuated from exterior mean-motion resonances
established by Fom b, even outside the planet's main chaotic zone.
The reasons for this are likely analogous to why
the solar system's Kirkwood gaps, located at Jupiter's interior
mean-motion resonances, are empty of asteroids \citep{wisdom82,wisdom83,wisdom85}.
Study of this phenomenon, which depends on the non-zero eccentricity
of the planet's orbit, is deferred to future work.

\begin{figure}
\centering
\scalebox{0.7}{\plotone{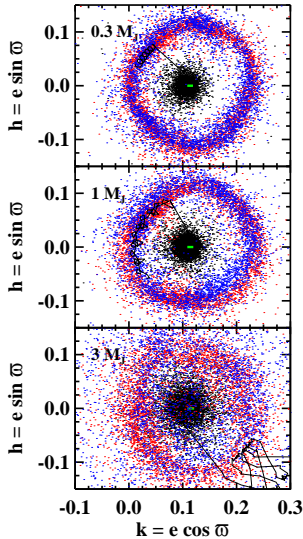}} 
\caption{Locations in complex eccentricity space of all parent bodies
  at $t=0$ (green bar) and those parent bodies that survive for
  $t=10^8\yr$ (black points).  The locus of survivors is not exactly
  centered on the locus of initial conditions because the survivors
  are all at large semimajor axes (smaller forced eccentricities), and
  because of inaccuracies in the L-L theory which was used to
  determine the initial conditions.  Surviving parent bodies remain
  approximately apsidally aligned with the planet's orbit, deviating
  by less than $\pm 15^{\circ}$ in most cases.  The deviations, i.e.,
  the dispersions in free eccentricity, increase with $M_{\rm
    pl}$. This same behavior was found by \citet{quillen06}.  However,
  contrary to that work, we find that the increased dispersion does
  not necessarily imply a more spatially diffuse inner edge to the
  belt; see Figure \ref{fig_taunorm} and \S\ref{sec_sharp}. Also shown
  are $\beta=0.1$ dust particles $10^4\yr$ after release (red points),
  and those same dust particles $10^5\yr$ after release (blue
  points). The trajectory of a typical dust particle is shown sampled
  every $10^3\yr$, starting from release. Note the large eccentricity
  variations for the 3 $M_{\rm J}$ model.}
\label{fig_hk}
\end{figure}

Figure \ref{fig_hk} plots the eccentricity vectors of the parent bodies
at the end of $10^8 \yr$ (black points).
While they remain clustered near their
initially purely forced values, there is a dispersion
that is not predicted by L-L: the parent bodies acquire free eccentricities
despite having none to start with. The more massive the planet, the greater
the free eccentricities that develop. This same behavior was found by
\citet{quillen06} and \citet{qf06}, who attributed it to
forcing by a mean-motion resonance just outside the chaotic zone.  

Figure \ref{fig_hk} also describes how dust particles (colored points)
born from parent bodies acquire free eccentricities. Initial
free eccentricity vectors are distributed roughly axisymmetrically
about the forced eccentricities. The initial free phase depends on
the orbital phase of release.  For example, a dust particle released
at the parent body's periastron gains a free eccentricity vector in
the $+\hat{\mathbf{k}}$ direction (total eccentricity increases), while
release at apastron yields a free eccentricity in the
$-\hat{\mathbf{k}}$ direction (total eccentricity decreases, at least
for $\beta$ not too large). Though these radiation-induced free eccentricities
of dust grains
are much larger than the free eccentricities acquired by parent bodies,
they do not necessarily lead to increased blurring of the belt inner edge,
because the semimajor axes of dust grains shift correspondingly outward
by radiation pressure as well (cf. equation \ref{eq_a}). 
We have verified
that the various $N_{\beta}$ surface density profiles for 
$M_{\rm pl} \leq 1 M_{\rm J}$ have comparably sharp inner edges for all $\beta$,
as gauged using our fractional width parameter $\delta$.
For further discussion of what influences the sharpness of the inner
edge, see \S\ref{sec_sharp}.

\begin{figure}
\centering
\scalebox{0.7}{\plotone{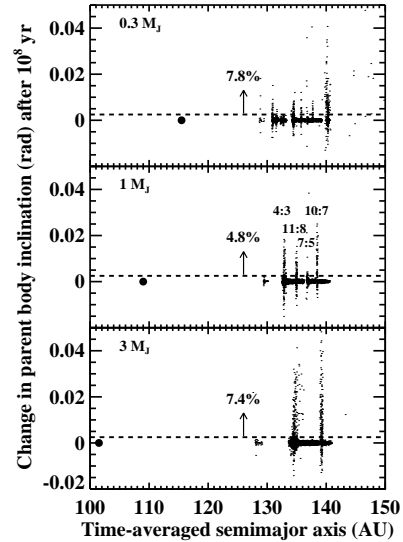}} 
\caption{Changes in orbital inclinations of surviving parent bodies,
  evaluated after $10^8\yr$. Solid circles denotes the planet.
  Only small percentages of stable parent
  bodies have their inclinations increased by the planet by more than
  $2.5 \times 10^{-3}$ rad (0.14 deg); these percentages are indicated
  in each panel. For comparison, the
  initial distribution of inclinations extends uniformly from 0 to $2.5 \times
  10^{-2}$ rad.  Those few objects that have their inclinations
  amplified are localized to mean-motion resonances, 
  labelled only for the middle panel.  }
\label{fig_inc}
\end{figure}

Finally, in Figure \ref{fig_inc} we examine how the orbital
inclinations of stable parent bodies change after $10^8
\yr$. Laplace-Lagrange predicts that the mutual inclination between
planet and particle is conserved, and indeed most inclinations are
little altered from their initial values (which span up to 0.025
rad). Fewer than 10\% of all surviving parent bodies have final
inclinations that exceed initial inclinations by more than 0.0025
rad. Excited bodies are localized to mean-motion
resonances. Even in the vicinity of a resonance, the fraction of
bodies that have their inclinations pumped is small. For example, in
our $1 M_{\rm J}$ model, only 10\% of all stable parent bodies having
time-averaged semimajor axes between 132 and 133.1 AU---particles
right at the inner
edge of the belt, in the vicinity of the 4:3 resonance---have
final inclinations exceeding initial inclinations by more than 0.0025 rad.
The near constancy of inclination should simplify future modelling
efforts: whatever vertical thickness of the belt
is desired to match the scattered light images can be
input into the dynamical model as an initial condition, in Step 1.

\subsection{Extensions and Refinements}\label{sec_extend}
In \S\ref{sec_sharp} we reconcile our results on
the sharpness of the inner belt edge with those of Q06.
In \S\ref{sec_tcol} we examine how robust our optical depth profiles
are to uncertainties in $t_{\rm col}$ and to our simplifying assumption
that the profile is adequately simulated by releasing grains from
parent bodies at a single time. In \S\ref{sec_resonance}
we experiment with different test particle
initial conditions to see whether they might yield superior
fits to the observations.

\subsubsection{Sharpness of Inner Belt Edge}\label{sec_sharp}
In Figure \ref{fig_tau_triple}, we found that the sharpness of the
belt's inner edge, as gauged by our measure $\delta$ (see
\S\ref{sec_res_tau}), hardly varied with $M_{\rm pl} \leq 1 M_{\rm
  J}$. This finding is seemingly at odds with that of Q06, who gauges
sharpness using the velocity dispersion of parent bodies at the
boundary of the chaotic zone, and who finds that it increases smoothly
as $\mu^{3/7}$. As a consequence of this relation, Q06 concludes that
$M_{\rm pl}$ cannot exceed $\sim$$7\times 10^{-5} M_{\ast} = 0.2
M_{\rm J}$ and still have the belt edge be as sharp as the observations
imply. Indeed we also found in Figure \ref{fig_hk} that the
velocity dispersion of parent bodies, as indicated by their
spread in free eccentricities, increased smoothly with $M_{\rm pl}$,
in apparent agreement with Q06. Yet the smoothly growing
velocity dispersion is not reflected
in the relative sharpness of our optical depth profiles across mass models.

How can this be? It might be thought that the discrepancy arises
because Q06's calculation of the velocity dispersion
pertains to parent bodies, while our calculation of $\tau_\perp$
involves dust. While as a point of principle our calculation would
be preferred because the observations are of dust and not of parent bodies,
this explanation does not get at the heart of the problem,
as we find the same invariant sharpness with planet mass
characterizing the surface
density profiles of our parent bodies.

The answer instead is that the sharpness of the inner edge
does not depend only on the velocity dispersion of particles
located strictly at the chaotic zone boundary. Particles located
at some radial distance from the boundary, further interior to the belt,
also contribute to the sharpness. That is because sharpness
is a relative quantity, measured relative to the maximum of $\tau_\perp$,
and this maximum
does not occur exactly at the chaotic zone boundary. Sharpness is appropriately
measured as a relative quantity,
since the observations are limited in dynamic range:
as Figure 3 of K05 indicates, only the maximum in surface brightness,
and values greater than $\sim$10\% of the maximum, are measurable.

\begin{figure}
\centering
\scalebox{0.7}{\plotone{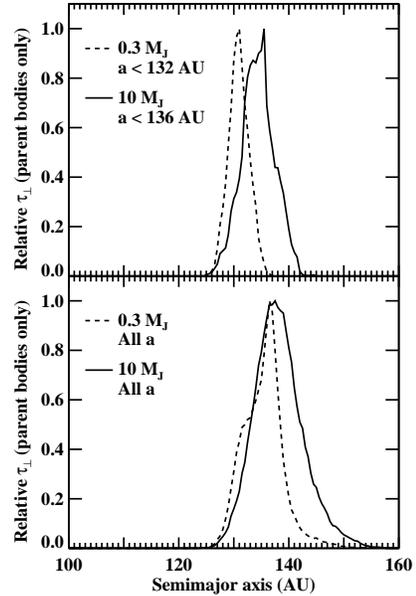}} 
\caption{Reconciling our findings on the sharpness of the belt inner
edge with those of Q06. {\it Top panel}: Surface density profiles
of stable parent bodies for two mass models, $0.3 M_{\rm J}$ and $10 M_{\rm J}$,
including only those bodies having the smallest semimajor axes
as indicated (cf. Figure \ref{fig_parent}). 
Only parent bodies are considered to compare fairly with Q06,
whose conclusions regarding inner edge sharpness pertain
to collisionless, radiation-free test particles.
In qualitative agreement with Q06, the $10 M_{\rm J}$
model yields a more diffuse inner edge, a consequence of that
model's larger velocity dispersion at the chaotic zone boundary
(cf. Figure \ref{fig_hk}).
{\it Bottom panel}:
Surface density profiles of the same two models with no restriction on
semimajor axes. Now the inner edges are of similar sharpness---in fact
the $10 M_{\rm J}$ profile appears slightly sharper than the $0.3 M_{\rm J}$
model---indicating
that sharpness is not uniquely related to edge velocity dispersion,
contrary to the implicit assumption of Q06.
}
\label{fig_taunorm}
\end{figure}

To illustrate our point, we plot in Figure \ref{fig_taunorm} the
surface densities of stable parent bodies for two mass models, $0.3
M_{\rm J}$ and $10 M_{\rm J}$. For the comparison with Q06 to be fair,
we must analyze only the parent bodies, since the conclusions of Q06
regarding inner edge sharpness pertain to collisionless,
radiation-free particles; in other words, the test particles of Q06
are our parent bodies. From Q06 we would expect that the $10 M_{\rm
  J}$ model produces an inner edge that is $(10/0.3)^{3/7} = 4.5
\times$ more diffuse, but the bottom panel of Figure \ref{fig_taunorm}
shows that this is not the case; in fact, if anything, the $0.3 M_{\rm
  J}$ profile appears more diffuse. In the top panel of Figure
\ref{fig_taunorm} we show the same two models except that we include
only particles having the smallest stable semimajor axes: $a < 132\AU$
for $M_{\rm pl} = 0.3 M_{\rm J}$ and $a< 136 \AU$ for $M_{\rm pl} = 10
M_{\rm J}$.  These profiles, which more strictly sample the chaotic
zone boundary, are more consistent with Q06---the $0.3 M_{\rm J}$
profile is sharper than the $10 M_{\rm J}$ profile. We conclude that
sharpness cannot be reliably computed without including contributions
from particles that are located some distance from the edge. Sharpness
cannot be calculated as if it were a local quantity specific to the
minimum stable semimajor axis; the shape of the observed surface
brightness profile reflects an amalgam of semimajor axes.

\subsubsection{Variations in $t_{\rm col}$ and Gradual Release of Grains}\label{sec_tcol}
\begin{figure}
\centering
\scalebox{0.7}{\plotone{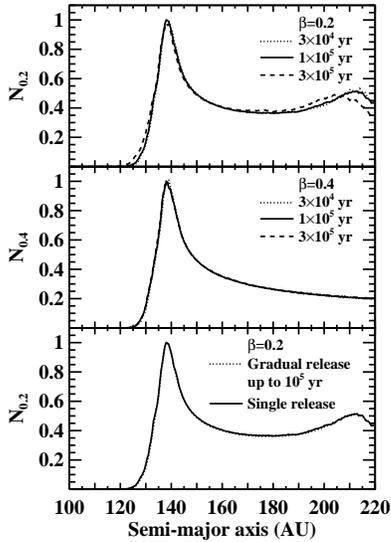}} 
\caption{{\it Top panel}: Surface density profiles for $\beta=0.2$
particles in our $1 M_{\rm J}$ model, as a function of $t_{\rm col}$,
whose value we only know to order of magnitude. Aside from a small
inward shift caused by PR drag for $t_{\rm col}=3 \times 10^5\yr$,
the profiles are not sensitive to our uncertainty in $t_{\rm col}$.
{\it Middle panel}: Same as top panel, except for $\beta=0.4$.
{\it Bottom panel}: Testing our simplifying assumption of a single
release time for dust grains. Allowing grains to be released
at ten uniformly spaced times between $t=0$ and $t = 10^5\yr$
produces no discernible difference from our standard model,
which assumes a single release time of $t=0$.
}
\label{fig_tcol}
\end{figure}

According to our standard procedure,
grain surface densities and optical depths from our
$\beta$-simulations are extracted after an integration time of $t_{\rm
  col}=10^5\yr$, a value inspired from our order-of-magnitude estimate
(\ref{eq_tcol}) of the collisional lifetime. In the top and middle
panels of Figure \ref{fig_tcol}, we demonstrate that our results are
not sensitive to uncertainties in $t_{\rm col}$. We present test results for
$\beta=0.2$ and $\beta=0.4$ since those cases contribute most to $\tau_\perp$
for our assumed Dohnanyi size distribution. Varying $t_{\rm col}$
from $3\times 10^4\yr$ to $3\times 10^5\yr$ produces practically
identical results for the surface
density of grains. Only for the $\beta=0.2$, $t_{\rm col} = 3 \times 10^5\yr$
run is there a slight $\sim$1 AU shift of the surface density profile,
a consequence of Poynting-Roberton drag. Such drag affects $\beta < 0.2$
grains even less. The highest $\beta$ grains are also relatively immune
from pericenter decay---see the middle panel of Figure \ref{fig_tcol}
for the case $\beta=0.4$---because of the large eccentricities and semimajor
axes induced by radiation pressure upon release \citep[][see also our
equation \ref{eq_tpr}]{ww50}.

We can also check whether our simple procedure of releasing grains at
a single time is sound. In reality, the belt at any given moment will
contain grains having a variety of ages, ranging from 0 (just released
grains) to $t_{\rm col}$ (grains just about to be shattered to sizes
small enough to be blown out by radiation pressure).  We better
simulate the gradual release of grains by adding together surface
density profiles computed from grains released at multiple times. We
divide the interval $t_{\rm col}$ into 10 release times, uniformly
spaced by $\Delta t = t_{\rm col}/10$, and generate separate
integrations for each time. That is, for the $i^{\rm th}$ release
time, we integrate the stable parent bodies (generated from Step 1)
forward with $\beta=0$ for $i\Delta t$, and then continue integrating
the released dust grains with $\beta\neq 0$ for $(10-i)\Delta t$.  The
resulting superposition of integrations, for $\beta=0.2$ and $t_{\rm
  col}=10^5\yr$, is shown in the bottom panel of Figure
\ref{fig_tcol}; it is nearly indistinguishable from our standard
result.  We have checked that ring shape is independent of grain age
$t$ for $1 / \Omega \ll t < t_{\rm col}$, because $t_{\rm col}$ is too
short a time for grains to evolve away from their birth orbits (say by
PR drag), and $1/\Omega$ is the timescale for grains to phase mix.

The experiments described in all three panels of Figure \ref{fig_tcol}
also show that our standard surface density and optical depth profiles do not
reflect dust features that vary with the orbital phase of the planet.
In varying $t_{\rm col}$ and superposing snapshots taken at different
times, we are extracting surface densities corresponding to random
orbital phases of the planet.

\subsubsection{Resonant Particles as Another Class of Stable Parent Body}\label{sec_resonance}
All our parent bodies are initialized with purely forced eccentric orbits,
as calculated using the Laplace-Lagrange secular theory. Their orbits
after $10^8\yr$ resemble their initial ones, with the addition of a small
free component.

\begin{figure}
\centering
\scalebox{0.7}{\plotone{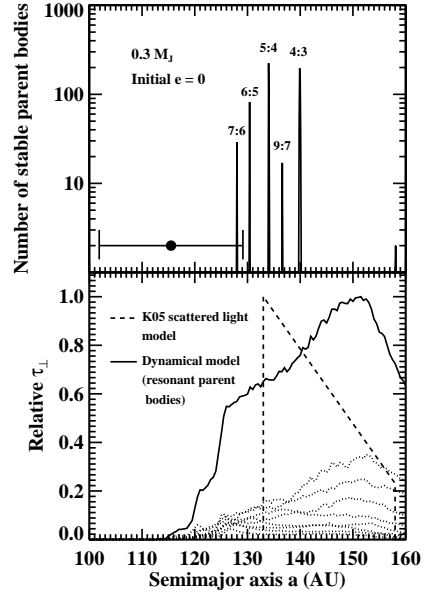}} 
\caption{Experimenting with initially zero eccentricities for parent bodies,
for the case $M_{\rm pl} = 0.3 M_{\rm J}$.
{\it Top panel}: Histogram of semimajor axes of parent bodies
that survive for $10^8\yr$, time-averaged over $10^5$ yr.
In stark contrast to our standard model (Figure \ref{fig_parent}),
survivors only inhabit mean-motion resonances, which afford
them protection from the close planetary encounters that would otherwise
result from the large eccentricities that develop secularly.
{\it Bottom panel}: Optical depth profile of dust generated
by resonant parent bodies. It is far too broad to agree
with the K05 model. We conclude that while resonant parent bodies
can exist in principle, their population in reality must be small
compared to that of parent bodies on nearly purely secularly forced orbits.
}
\label{fig_resgrand}
\end{figure}

Here we try a different set of initial conditions:
osculating $e=0$. According to L-L, this corresponds to
assigning particles free eccentricities equal in magnitude to 
their forced eccentricities (see, e.g., \citealt{md00}).
The top panel of Figure \ref{fig_resgrand} documents
the resultant time-averaged semimajor axes of stable parent bodies (those
that survive for $10^8\yr$), for the case
$M_{\rm pl}=0.3 M_{\rm J}$. All other initial conditions apart from the test
particles' eccentricities are specified the same way as for our standard
$0.3 M_{\rm J}$ model.
Remarkably, comparing the top panels of Figures \ref{fig_resgrand} and
\ref{fig_parent},
we find the distributions of stable semimajor axes are nearly
complementary. Instead of being cleared out of mean-motion
resonances, stable parent bodies inhabit them exclusively
when initial eccentricities equal zero.

The resonances---which include the 7:6, 6:5, 5:4, 9:7, and 4:3, and
which are of eccentricity-type---protect the particles from close encounters
with the planet.
Qualitatively, the eccentricities and apsidal angles behave as L-L
predicts: while $e$ cycles from 0 through
$\max(e) = 2e_{\rm forced} \approx 0.22$ back to 0, $\pomega$
regresses from $\pi/2$ to $-\pi/2$ (the evolution of $\pomega$
is discontinuous since $e$ passes through 0). The large maximum eccentricities,
which result because the free eccentricities are of the same magnitude
as the forced eccentricities,
put particles in danger of close planetary encounters,
especially when $e=\max(e)$ and $\pomega=0$.
Under these conditions,
for a semimajor axis of, say, $a = 128\AU$, the particle's pericenter 
encroaches within
$\sim$$2 \AU \approx 0.5 R_{\rm H}$ of the planet's pericenter.

\begin{figure}
\centering
\scalebox{0.8}{\plotone{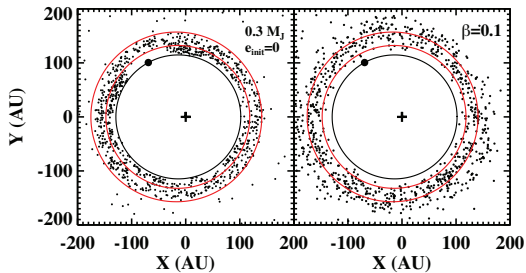}}
\caption{Same as Figure \ref{fig_snap100}, except
for resonant parent bodies. The left panel shows how
resonant parent bodies avoid close encounters with the planet.
The entire pattern corotates with the planet.
Dust grains released from these resonant parent bodies have large
eccentricities and yield an optical depth profile too broad
to match that of K05; see Figure \ref{fig_resgrand}.
}
\label{fig_snapres}
\end{figure}

However, thanks to resonance, conjunctions occur only at special
orbital phases and close encounters do not occur.
For the circumstances just described, during the phase that
the particle attains maximum eccentricity,
we have observed in our numerical integrations that
conjunctions never occur at periastron. Because of the 7:6 resonance,
they occur instead 180$^{\circ}$ degrees away, at apoapse,
when the bodies are well separated by about 27 AU. Figure \ref{fig_snapres} provides
a snapshot of stable resonant parent bodies, showing how they
avoid approaching the planet.

Can such resonant parent bodies be present in Fomalhaut's belt?  Not
in significant numbers compared to the non-resonant population.  The
resonant bodies develop more eccentric orbits, and consequently
the dust they produce is more spatially extended. From the bottom
panel of Figure \ref{fig_resgrand} (see also the right-hand panel of
Figure \ref{fig_snapres}), it is clear that the optical depth profile
of dust released from resonant parent bodies is far too broad to match
that of K05.  (Our procedure of calculating optical depths by
smoothing particles over their orbits is not correct for resonant
objects, since the smoothing ignores their special orbital phase
relationships with the planet. However the error accrued is small,
since $\tau_\perp$ is dominated by dust particles having $\beta
\gtrsim 0.1$. Such dust particles, upon release, have their semimajor
axes increased by $\gtrsim 10$\% by radiation pressure, and are thus
removed from the resonances inhabited by their parents.)

In \S\ref{sec_ori} we discuss how the parent bodies might have come
to occupy nearly purely secularly forced orbits and to avoid the
resonant orbits.

\section{SUMMARY AND DISCUSSION}\label{sec_sum}

We review our main results in \S\ref{sec_summ}; sketch
the effects of other planets apart from Fom b in \S\ref{sec_other};
and discuss possible origins of Fom b and the belt
in \S\ref{sec_ori}.

\subsection{Summary}\label{sec_summ}
Fomalhaut b is the first extrasolar planet candidate to be directly
imaged at visible wavelengths and to have its orbital motion around
its host star measured.
Surprises have
been immediate: Fomalhaut b has an unusually large orbital radius of
more than 110 AU \citep[cf.][]{laf08}, and a visual (0.45--0.7 $\mu$m)
luminosity that is not only 1--2 orders of magnitude greater than
atmospheric models anticipated, but also time variable. Nevertheless,
a chain of arguments based on comparing the observed photometry with
model exoplanet atmospheres leads Kalas et al. (2008, K08) to infer
that the mass of Fomalhaut b must be less than about $3 M_{\rm J}$.

The Fomalhaut system is all the more remarkable for offering a means
independent of model spectra to get at Fom b's mass: the star is
encircled by a belt of dust whose geometry is, in principle, sensitive
to the mass and orbit of Fom b. At a system age of $\sim$200 Myr,
detritus from the formation of the Fomalhaut planetary system still
remains. If Fom b is the sole sculptor of this debris---but see
\S\ref{sec_intro} and \S\ref{sec_other} for reasons that it might not
be---then the observed intrinsic ellipticity of the ring would owe its
origin to secular forcing by Fom b, which itself would reside on an
apsidally aligned and similarly eccentric orbit to the belt's
\citep{wyatt99,quillen06}.  Under the single planet assumption,
another feature of the belt influenced by Fom b would be its inner
edge. The observed sharpness with which the belt truncates would reflect
the sharp divide between stability and chaos at the boundary of the
planet's chaotic zone, inside of which first-order mean-motion
resonances overlap and particle orbits are short-lived
\citep{w80,quillen06}.  Particles inside the zone quickly evolve onto
planet-crossing orbits and thereafter are perturbed onto escape
trajectories.\footnote{ We estimate that planet crossing takes
  $\sim$$10^5 (10^{-3}/\mu)^{4/7}$ yr and ejection takes $\sim$$10^7
  (10^{-3}/\mu)^{2}$ yr, for particles halfway between the planet and
  the edge of the chaotic zone.}  Identifying the belt inner edge with
the chaotic zone boundary relates the distance between the planet and
the belt edge to the planet-star mass ratio.

Based on these and other theoretical ideas, we have built a realistic
dynamical model of the Fomalhaut planet-belt system, under the single
planet assumption.  Our goal is to
calculate the spatial distribution of dust generated from the
collisional comminution of larger parent bodies, and to compare our
dust maps with the HST scattered light observations.  The model begins
by using numerical integrations to establish an annulus of parent
bodies---a.k.a., the ``birth ring'' \citep{sc06}---that is stable for
100 Myr against perturbations by Fom b.  Dust particles are released
from these dynamically stable parents, and their trajectories
followed, taking additional account of stellar radiation forces, for a
dust collisional lifetime of $t_{\rm col} = 0.1$ Myr.  The orbits of
dust particles at the end of a $t_{\rm col}$-long integration are
assumed to represent well those of actual dust particles, which in
reality have a range of ages extending up to $\sim$$t_{\rm col}$. We
have subjected this assumption to a few tests and found it to
hold. Final dust particle orbits, computed for a range of radiation
$\beta$'s (force ratio between stellar radiation pressure and stellar
gravity), are converted into orbit-averaged maps of relative surface
density (number of grains per unit face-on area of the belt).  A
\citet{dohn69} grain size distribution, appropriate for a quasi-steady
collisional cascade, is assumed to hold in the birth ring, where the
dust surface density is highest and collision rate is greatest.  This
assumption determines how we weight and add the surface density maps
to produce a map of (relative) vertical optical depth.  That optical
depth map---or rather its azimuthally averaged version, the variation
of optical depth with semimajor axis---is compared with the optical
depth profile of the \citet{kalas05} scattered light model, which
itself represents an idealized and approximate fit to the HST images.

A conservative result of our dynamical model is that $M_{\rm pl} < 3 M_{\rm J}$.
This conclusion, that Fom b must be of planetary mass, is entirely
independent of Fom b's photometry---and its uncertain interpretation.
Our result stems from two simple and robust trends, neither of which
involve inner edge sharpness, in contrast to Q06. First, as $M_{\rm
  pl}$ increases, dust particles are increasingly perturbed by the
planet onto more eccentric orbits, rendering the dynamical profiles
too broad compared to the scattered light profile. A mass of $10
M_{\rm J}$ yields a belt that is about twice as bright between 150 and
160 AU as the observations allow, while lower masses $M_{\rm pl} < 3
M_{\rm J}$ give optical depth profiles that we feel agree adequately
well with the \citet{kalas05} scattered light model.  Second, given
the observed location of the inner edge of the belt, larger mass
planets have necessarily smaller orbits located farther interior to
the belt, and smaller orbits may be incompatible with the observed
stellocentric distance of Fom b.  A mass of $10 M_{\rm J}$ requires an
orbit whose apocentric distance is 107 AU, falling well short of our
estimated de-projected distance for Fom b of 119 AU. By contrast, for
$M_{\rm pl} \leq 1 M_{\rm J}$, the model apocentric distances are
$\geq 122$ AU.  While $3 M_{\rm J}$ yields an orbit whose apocenter
lies at 115 AU and is nominally incompatible, uncertainties in the
observed de-projected distance, probably amounting to a few AU,
preclude us from ruling out this mass.  Thus, erring on the safe side,
we conclude that $M_{\rm pl} < 3 M_{\rm J}$.

Our findings
agree in broad outline with those of \citet{quillen06}, whose
prediction that the belt inner edge be located at the boundary of the
planet's chaotic zone appears vindicated by the discovery of Fom b (modulo
the nominal finding that belt and planet orbits are apsidally
misaligned---whether the misalignment is real will have to await
multi-epoch astrometry of Fom b).
We diverge from \citet{quillen06} in how we determine Fom b's
mass.  Among the improvements we make are: (1) we draw a clear
distinction between unobservable parent bodies and observable dust
grains, and rely on the latter when comparing with the HST scattered
light observations; (2) we include the effect of stellar radiation
pressure, significant for dust grains; (3) parent bodies are screened
for dynamical stability over the age of the system; and (4)
grain-grain collisions are recognized as destructive, and therefore
the duration of each of our dust particle integrations is necessarily
limited by the collision time.  Our upper mass limit of $3
M_{\rm J}$ follows, in part, from comparing theoretical and observed
optical depth profiles of dust, computed globally over all space, and
from noting how steep those profiles are outside of
the location of peak optical depth.  By contrast, the upper mass limit
derived by \citet{quillen06}, $\sim$$0.2 M_{\rm J}$, follows from an
analysis of the radiation-free dynamics of collisionless
particles---essentially, the dynamics of parent bodies---local to the
chaotic zone boundary, and from the assumption that the local velocity
dispersion of such bodies determines the sharpness of the inner belt
edge. Our calculation of the upper mass limit is preferred because the
HST observations are of dust, not of parent bodies, and also because
we have shown that edge sharpness is better computed using a global
model such as ours.

Three Jupiter masses for the mass of Fom b is an upper limit in still
another sense: our results are based on the assumption that Fom b alone sculpts
the belt. Roughly speaking, the more planets that are present, the
more orbits are chaotic, and the more
effectively small bodies and dust are gravitationally scoured.
Thus the observed inner edge of Fomalhaut's belt
is compatible
with a mass for Fom b
lower than that computed under
the single planet assumption.
It is heartening to see that our preferred mass range
of $M_{\rm pl} < 3 M_{\rm J}$ supports
that inferred from the spectral models.  

Perhaps the biggest deficiency of our model lies in our crude
treatment of grain collisions. In setting a dust particle's initial
position and velocity equal to that of its parent body, we ignore
collisional dissipation and redirection of orbital kinetic energy.  We
also neglect the fact that grinding the largest parent bodies down to
dust requires multiple collisions, and that radiation effects can
start manifesting in the middle of the collisional cascade. For
example, a particle for which radiation effects are significant, say
which has $\beta=0.4$, can be born from a parent body for which
radiation effects were also significant, say which had
$\beta=0.2$. Still, our simple prescription of releasing dust
grains having $\beta > 0$ from bodies having $\beta = 0$ is not
without justification. Collisional energy dissipation should, on average,
dampen free eccentricities but not forced eccentricities (see \S\ref{sec_ori});
thus the mean elliptical shape of the belt is expected to be preserved.
Grain ejection velocities relative to the parent
are distributed isotropically and should therefore
not bias our results, though they will produce larger free
eccentricities than our model predicts. Larger free eccentricities
will result in smoother optical depth profiles: a blurring of the
belt (see \S\ref{sec_ori} for further discussion of free eccentricities).
Finally, radiation effects are predominantly felt over only
the last decade in grain size above the blow-out size, i.e., only after
the penultimate collision just prior to the final collision resulting
in blow-out, for collisions that are strongly disruptive.
The next generation of models should test these assertions,
in addition to considering qualitatively different physics
(e.g., Yarkovsky drag,
and gas-particle interactions.\footnote{Fomalhaut's closest analogue
may be AU Mic, insofar as both have birth ring morphologies and similar
optical depths \citep{sc06}. Gas has not been detected in AU Mic
(Brandeker et al., 2008 Spitzer Science Conference Poster \#81).})

\subsection{Other Planets In Addition to Fom b}\label{sec_other}

As discussed in \S\ref{sec_intro}, the observed space velocity of Fom
b is nominally inconsistent with its orbit being apsidally aligned
with that of the belt, contradicting a basic prediction of the hypothesis
that Fom b is solely responsible for the morphology of
Fomalhaut's debris disk. The final word on orbits must await more
epochs of astrometry and more realistic assessments of systematic
uncertainties (in, e.g., frame registration between
epochs).
Even if the apsidal misalignment proves
real, and even if follow-up observations constrain Fom b's mass to
be less than that required to make a significant contribution
to shaping the belt,
ideas of chaotic zone clearing and secular forcing
of eccentricity may still be relevant for the Fomalhaut system,
provided there are perturbers in addition to Fom b.

We have already argued that multiple planets are compatible with
the inner edge to Fomalhaut's debris disk; more planets help eject more material
from inside the hole. Multiple planets are also compatible
with the observed belt eccentricity. In Laplace-Lagrange theory, the
observed forced eccentricity vector of a belt particle equals the
vector sum of $n$ eccentricity vectors forced by $n$ planets. Because
the individual eccentricity vectors precess at frequencies that depend
only on the fixed masses and semimajor axes of the planets (these are
the fixed eigenfrequencies of the linear theory), and because belt
particles having similar semimajor axes have similar vector
decompositions, we expect the forced eccentricity vectors of belt
members to remain similar to one another over time. Thus a narrow belt
can maintain a global mean eccentricity in the presence of multiple
planets, though that mean eccentricity will oscillate with time.  All
these considerations can be accommodated as necessary into our
modelling procedure.

If future astrometry confirms
a significant apsidal misalignment between planet and belt,
then at least one other, as yet
unseen planet would be implicated.  In that case, because the belt's
forced eccentricity is a vector sum, Fom b's eccentricity
could either be lower or higher than the eccentricity we have
calculated, depending on the apsidal orientation(s) of the other
planetary orbit(s).

Intriguingly, over its three-year mission, the
Hipparcos satellite observed Fomalhaut to have an ``anomalous'' proper
acceleration of 6.6 milliarcsec/yr$^2$,
of marginal (about 2$\sigma$) significance. If real,
such a quasi-steady
acceleration might be caused by a companion whose orbital period is
longer than $\sim$3 years, or equivalently whose stellocentric
distance $r \gtrsim 3 \AU$.  Equating the measured acceleration with
$(GM/r^2)/d$, where $M$ is the perturbing mass and $d= 7.7$ pc is the
distance to Fomalhaut, we see that Fomalhaut might also be harboring a
$\sim$$30 M_{\rm J}$ brown dwarf at a distance $r \sim 5
\AU$. (Larger $r$ implies larger $M$ and such solutions are probably
ruled out by observation.  Clearly Fom b cannot be responsible for the
Hipparcos acceleration.)  Compared against the influence of Fom b as
we have computed it in this paper, such a brown dwarf would contribute
more than 10\% to the forced eccentricity of a belt particle, if the
brown dwarf's eccentricity $\gtrsim 0.2$.

The above lines of evidence for additional planets perturbing the belt
are tenuous. Given the observed proximity of Fom b to the belt, the
opposing case can be made that Fom b dominates the forced
eccentricities of belt particles.  A possible analogy would be Neptune
and the Kuiper belt. Despite the existence of as many as four giant
planets in our solar system, the forced inclination and eccentricity
vectors of Kuiper belt objects are largely determined by the nearest
planet, Neptune (see, e.g., \citealt{cc08}).  If Fom b's
orbit is apsidally aligned with the belt, then Fom b's observed
deprojected stellocentric distance of 119 AU (Section 3.2.3)
and current true anomaly of about 109 degrees would
imply a semimajor axis of 114--115 AU.
Given such a semimajor axis, our model would
predict $M_{\rm pl} \approx 0.5 M_{\rm J}$ (Table 1).

\subsection{Parent Body and Planet Origins}\label{sec_ori}

Though the direct detection of parent bodies is beyond the reach of
current observations, our study has provided some evidence that they
reside mostly on nearly purely secularly forced orbits with small free
eccentricities. In principle, they do not have to; we found by
experimentation in \S\ref{sec_resonance}
that large free eccentricities are also compatible
with long-term stability if the particles are protected by mean-motion
resonances.

How did the parent bodies choose one class of stable orbit over the other? 
The answer, as suggested also by \citet{qf06}, likely involves
collisional dissipation.  Collisions dissipate random orbital motions
and compel planetesimals to conform towards closed, non-intersecting
orbits viewed in the frame rotating with the perturbation potential
\citep[e.g.,][section 5.4]{pac77,gt82}. If that perturbation potential
arises from Fom b,
the special closed orbits include the secularly forced orbits that we
have been highlighting throughout our study, but they do not include
the resonant orbits that emerged from our experiment.  Collisional
dissipation and relaxation onto closed orbits occur across the entire
collisional cascade, up to the largest parent bodies (which by
definition collide once over the system age).

We do not expect the destructive nature of collisions to qualitatively
alter this picture, since what is important here is the dissipation of
random kinetic energy, and that occurs whether or not collisions are
destructive.  Post-collision fragments will have free eccentricities
that are small compared to forced eccentricities, insofar as
post-collision fragment velocities (measured relative to the center of
mass) are small compared to $e_{\rm forced} \Omega_R R \sim 400$ m/s.
At least in the collisional genesis of the Eunomia and Koronis
asteroid families in our solar system's main belt, ejection
velocities of the largest post-collision remnants range 
from 4 to 90 m/s \citep{michel}.

It is also possible that relaxation occurred while the parent bodies were
forming, when collisions were gentler and agglomerative.
The process of relaxation onto forced orbits can be explored
using fast numerical simulation techniques
for inelastically colliding, indestructible particles \citep{lc07}; in fact,
we have started to run such simulations and clearly observe relaxation.

What are the origins of Fom b and the belt?  It seems likely that the
belt is what remains of the original disk material that went into
building Fom b.  If we take the minimum parent body mass of $3
M_{\oplus}$ (estimated in \S\ref{sec_totparent}) and augment it by a
factor of $10^2$ to bring it to cosmic composition, then the minimum
primordial mass for the belt is $\sim$$1 M_{\rm J}$.  This is
comparable to our upper mass limit for Fom b. A working
hypothesis is that Fom b accreted {\it in situ} from a primordial disk
of gas and dust; that the hydrogen gas of the original disk has either
accreted into planets or photoevaporated; and that today the remaining
solids in the belt are grinding down to dust, the in-plane velocity
dispersion of parent bodies excited so strongly by Fom b, and possibly
other perturbers, that
collisions are destructive rather than agglomerative.  The last tenet
is supported by our Figure \ref{fig_hk}, which shows free eccentricity
dispersions that imply relative parent body velocities upwards of
$\sim$100 m/s (see also our \S\ref{sec_destruc}).

\acknowledgements This work was supported by NSF grant AST-0507805. E.K.
acknowledges support from a Berkeley Fellowship. We
thank Mike Fitzgerald, Yoram Lithwick, 
Eric Mamajek,
Christian Marois,
Norm Murray,
Karl Stapelfeldt, and Mark Wyatt for discussions. An anonymous referee
rapidly provided a thoughtful report, at the behest of ApJ Scientific
Editor Fred Rasio, for which we are grateful.

\bibliographystyle{apj}

\begin{deluxetable}{ccccccccccccc}
\tablecaption{Possible Properties of Fom b and Numbers of Surviving Belt Particles}
\tabletypesize{\tiny}
\tablewidth{0pt}
\tablehead{
\multicolumn{13}{c}{}\\
$M_{\rm pl}$ & $a_{\rm pl}$ & $e_{\rm pl}$ & $N_{\rm tp}$ & $N_{\rm tp}$ & $N_{\rm tp}$ & $N_{\rm tp}$ & $N_{\rm tp}$ & $N_{\rm tp}$ & $N_{\rm tp}$ & $N_{\rm tp}$ & $N_{\rm tp}$ &  $N_{\rm tp}$ \\
$(M_{\rm J})$ & (AU) &   & $(t=0)$ & $(t=10^8\yr)$ & $(0)$$^a$ & ($0.00625$) & ($0.0125$) & ($0.025$) & ($0.05$) & $(0.1)$ & ($0.2$) & ($0.4$) \\
}
\startdata
0.1$^b$  & 120  & 0.115  & $10^4$  & 5798 & 5798 & 5796  & 5798 & 5798 & 5798 & 5798 & 5798 & 5727 \\

0.3  & 115.5  & 0.118 & $10^4$ & 4471 & 4471 & 4467 & 4471 & 4471 & 4471 & 4471 & 4470 & 4421 \\

1 & 109 & 0.123 & $10^4$ & 4191 & 4191 & 4191 & 4191 & 4191 & 4191 & 4191 & 4191 & 4187 \\

3  & 101.5  & 0.130  & $10^4$ & 4466 & 4466 & 4456  & 4454 & 4465 & 4465 & 4466 & 4466 & 4456 \\

10$^c$ & 94  & 0.138 & $10^4$ & 1548 & 1548 & 1546 & 1539 & 1433 & 1033 & 1460 & 1528 & 1548 \\
\enddata

\tablenotetext{a}{The pure number in parentheses for this column and subsequent columns equals $\beta$. All $\beta$-simulations tabulated here are integrated for $t_{\rm col} = 10^5\yr$, following the $10^8$-yr-long parent body integration.}

\tablenotetext{b}{For our 0.1 $M_{\rm J}$ run, initial parent body semimajor axes extend from 125 AU to 140 AU.}

\tablenotetext{c}{The $10 M_{\rm J}$ run is not favored since it produces a vertical optical depth profile that does not match that of the K05 scattered light model.}
\label{tab_one}
\end{deluxetable}

\end{document}